\documentclass[11pt, reqno]{amsart}
\usepackage{fullpage}
\usepackage[OT1]{fontenc}
\usepackage{amsthm,amsmath,amsfonts,amssymb,amsaddr}
\usepackage{hyperref}
\usepackage{xcolor}
\hypersetup{
    colorlinks=true,
    linkcolor=purple,    
    urlcolor=blue,
    citecolor=blue
}

\usepackage{mathrsfs}
\usepackage{bm}
\usepackage{fancybox}
\usepackage{graphicx}
\usepackage{xcolor}
\usepackage{booktabs}
\usepackage{multirow}
\usepackage{subcaption}
\usepackage[round]{natbib}
\usepackage{algorithm}
\usepackage[noend]{algpseudocode}

\newtheorem{theorem}{Theorem}

\newtheorem{lemma}{Lemma}

\theoremstyle{definition}

\def\T{{ \mathrm{\scriptscriptstyle T} }}

\newcommand{\given}{\mid}
\newcommand{\biggiven}{\,\middle|\,}
\newcommand{\E}{\mathbb{E}}

\newcommand{\GP}{\mathcal{GP}}
\newcommand{\EF}{\mathsf{EF}}
\newcommand{\DY}{\mathsf{DY}}
\newcommand{\Norm}{\mathsf{N}}
\newcommand{\MNorm}{\mathsf{MN}}
\newcommand{\IG}{\mathsf{IG}}
\newcommand{\IW}{\mathsf{IW}}
\newcommand{\matrixT}{\mathsf{MT}}
\newcommand{\GCM}{\mathsf{GCM}}
\newcommand{\GCMc}{\mathsf{GCM_c}}
\newcommand{\thetasp}{{\theta_{\text{sp}}}}
\newcommand{\calD}{\mathcal{D}}
\newcommand{\calL}{\mathcal{L}}
\newcommand{\calS}{\mathcal{S}}
\newcommand{\calT}{\mathcal{T}}
\newcommand{\calM}{\mathcal{M}}

\newcommand{\tr}{\mathsf{tr}}

\title[Bayesian inference for spatial-temporal non-Gaussian data]{Bayesian inference for spatial-temporal non-Gaussian data using predictive stacking}

\author{Soumyakanti Pan$^{\star}$, Lu Zhang$^{\dagger}$, Jonathan R. Bradley$^{\ddagger}$, Sudipto Banerjee$^{\star}$}
\address{$^\star$Department of Biostatistics, University of California Los Angeles\\$^\dagger$Division of Biostatistics, Department of Population and Public Health Sciences,\\University of Southern California\\$^\ddagger$Department of Statistics, Florida State University}

\date{\today}

\begin{document}
\begin{abstract}
    Analysing non-Gaussian spatial-temporal data requires introducing spatial as well as temporal dependence in generalised linear models through the link function of an exponential family distribution. Unlike in Gaussian likelihoods, inference is considerably encumbered by the inability to analytically integrate out the random effects and reduce the dimension of the parameter space. Iterative estimation algorithms struggle to converge due to the presence of weakly identified parameters. We devise Bayesian inference using predictive stacking that assimilates inference from analytically tractable conditional posterior distributions. We achieve this by expanding upon the Diaconis-Ylvisaker family of conjugate priors and exploiting generalised conjugate multivariate ($\GCM$) distribution theory for exponential families, which enables exact sampling from analytically available posterior distributions conditional upon some process parameters. Subsequently, we assimilate inference over a range of values of these parameters using Bayesian predictive stacking. We evaluate inferential performance on simulated data, compare with full Bayesian inference using Markov chain Monte Carlo (MCMC) and apply our method to analyse spatially-temporally referenced avian count data from the North American Breeding Bird Survey database.
    
    \smallskip
    \noindent \textbf{Keywords.} Spatial GLM; model combination; Spatially-temporally varying coefficients.
\end{abstract}
\maketitle

\section{Introduction}
Statistical modelling for spatially oriented non-Gaussian outcomes plays a crucial role in various scientific applications \citep[see, for example,][]{deoliveira1997tgrf, diggle_geostat_1998, heagerty1998composite, de2000bayesian, zhang2021tractable, saha_datta_banerjee_2022}. For example, climate scientists record daily or monthly binary variables at spatial locations indicating whether or not rainfall was measurable; ecologists analyse the temporal evolution of species counts at locations; and economists study the spatial distribution of the number of insurance claims over time. 

For our purposes, we consider a spatial-temporal process as an uncountable set of random variables, say $\{z(\ell): \ell\in \calD\}$, which is endowed with a probability law specifying the joint distribution for any finite sample of locations in $\calD\subset \mathbb{R}^d$ (with $d=2$ or $d=3$). In spatial-temporal settings $\calD = \calS\times \calT$, where $\calS \subset \mathbb{R}^d$  and $\calT \subset [0,\infty)$ are space and time domains, respectively, and $\ell = (s,t)$ is a space-time coordinate with $s \in \calS$ and $t\in \calT$ \citep[see, e.g.,][]{gnei10}. 

Following \cite{diggle_geostat_1998}, we introduce a spatial-temporal stochastic processes for non-Gaussian data. Let $y(\ell)$ be the outcome at $\ell$ endowed with a probability law from the natural exponential family, which we denote by 
\begin{equation}\label{eq:NEF}
    y(\ell) \sim \EF(x(\ell)^\T \beta  + z(\ell); b, \psi_y)
\end{equation}
for some positive parameter $b > 0$ and the unit log partition function $\psi_y$ (Section~\ref{subsec:dy}). Fixed effects and spatial dependence, e.g., $x(\ell)^{\T}\beta + z(\ell)$, are introduced in the natural parameter, where $x(\ell)$ is a $p \times 1$ vector of predictors referenced with respect to $\ell$, $\beta$ is a $p \times 1$ vector of slopes, $z(\ell)$ is a zero-centred spatial-temporal Gaussian process on $\mathbb{R}^d$ specified by a scale parameter $\sigma_z$ and a spatial-temporal correlation function $R(\cdot, \cdot ; \thetasp)$ with $\thetasp$ consisting of spatial-temporal decay and smoothness parameters. This structure is embodied by spatial-temporal generalised mixed-effect models \citep{glmm_mcculloch, hughes_haran_2013}. Bayesian inference for \eqref{eq:NEF} is appealing as it offers fully probabilistic inference for the latent process. However, the presence of $z(\ell)$ presents challenges for non-Gaussian families as we cannot integrate out $z$, which begets a high-dimensional parameter space. Iterative algorithms such as Markov Chain Monte Carlo (MCMC) attempt to sample from the posterior distribution, but are often encumbered by high auto-correlations and weakly identified parameters $\thetasp$. Section~1.3 in \cite{haran2011} overviews MCMC algorithms for estimating such models. 

This article departs from the customary focus on improving computational algorithms for spatial generalised linear models (GLMs). Instead, we devise Bayesian predictive stacking to conduct Bayesian inference for fixed effects and the latent spatial-temporal process by avoiding MCMC or other iterative algorithms for spatial-temporal non-Gaussian data. Stacking \citep{wolpert1992stacked, breiman1996stacked, clyde2013bayesian} is a model averaging procedure that is widely used in machine learning and has been shown \citep[see, e.g.,][]{le2017bayes, yao2018using, yao2020stacking, yao21_duplicate, zhang2025jasa} to be an effective alternative to traditional Bayesian model averaging \citep{madigan1996bayesian, hoeting1999bma}. We divide our model parameters into two sets $\theta_1 = \{\beta, z, \sigma\}$ and $\theta_2 = \{\thetasp, \cdot\}$, where $\sigma$ denotes a collection of scale parameters associated with $\beta$ and $z$, and $\cdot$ represents other auxiliary model parameters. We use conjugate distribution theory so that $p(\theta_1 \given y, \theta_2)$ is available in closed form. We are primarily concerned with Bayesian inference on $\theta_1$ averaging out $\theta_2$. Therefore, our desired posterior distribution is $p(\theta_1 \given y) = \int p(\theta_1\given y, \theta_2)p(\theta_2\given y) d\theta_2$. The key bottleneck here is that $p(\theta_2\given y)$ is intractable. This further precludes empirical Bayes strategies, as evaluating $p(\theta_2 \given y)$ requires marginalization over $\theta_1$, which entails a high-dimensional numerical integration and is itself computationally expensive. Therefore, we reformulate our inference problem by writing $p(\theta_1 \given y) = \int p(\theta_1\given y, \theta_2)p(\theta_2\given y) d\theta_2 \approx \sum_{g=1}^G w_g p(\theta_1\given y, \theta_{2g})$, where the collection of weights $w_g$ replaces $p(\theta_2\given y)$. A key distinction from quadrature based methods \citep[e.g., INLA,][]{inla2009} is that we do not attempt to approximate $p(\theta_2\given y)$. Instead, we find $w_g$ using convex optimisation with scoring rules. Once the optimal weights, say $\hat{w}_g$, are computed, posterior inference for quantities of interest proceeds from the ``stacked posterior'' $\tilde{p}( \cdot \given y) = \sum_{g = 1}^G \hat{w}_g p(\cdot \given y, \theta_{2g})$.

We develop and investigate Bayesian predictive stacking for spatial-temporal GLMs \citep[thereby extending][to spatial-temporal non-Gaussian data]{zhang2025jasa}. This requires hierarchical models to produce analytically accessible posterior distributions for $\theta_1$ subject to fixing weakly identified parameters $\theta_2$ and some hyperparameters to draw exact posterior samples for any fixed values of these parameters. Here, we avail of a new class of analytically accessible generalised conjugate multivariate distributions ($\GCM$) for spatial models \citep{bradley2023lgp}  by extending the Diaconis-Ylvisaker family of conjugate priors for exponential families \citep{DY79}. We significantly expand on this framework by considering spatial-temporal processes and further enrich \eqref{eq:NEF} by introducing spatially-temporally varying regression coefficients \citep{gelfand2003svc} to capture the impact of predictors over space and time. Bayesian modelling is especially appealing here, as it offers inference on processes that are completely unobserved. 

The literature on spatially-temporally varying coefficient models is rather sparse and has remained purely notional for exponential families, perhaps because they require extensive tuning of MCMC or other iterative algorithms that struggle with weakly identified parameters and high-dimensional random effects. Our subsequent development offers, as we demonstrate below, a fast and effective inferential tool for analysing spatial-temporal count or binary data. The analytical tractability we seek will forego inference on some weakly identifiable parameters ($\theta_2$ above), but the key inferential objectives concerning the natural mean function (e.g., regression coefficients), the spatial-temporal process and the variance components are retained. We show that the inference obtained on these parameters is practically indistinguishable from full inference using MCMC.

After briefly discussing some necessary distributions in Section~\ref{sec:conjugate}, Section~\ref{sec:bayeshier} introduces our hierarchical spatially-temporally varying regression model and derives analytical posterior distributions. Section~\ref{sec:stacking} develops predictive stacking with novel results on posterior sampling and predictive inference. Sections~\ref{sec:simulation}~and~\ref{sec:data_analysis} present simulation experiments demonstrating the effectiveness of our method and analyse a spatial-temporal bird counts data. Section~\ref{sec:discussion} concludes with a brief discussion.

\section{Conjugate priors for exponential family}\label{sec:conjugate}
\subsection{The Diaconis-Ylvisaker distribution}\label{subsec:dy}
Let $Y$ be distributed from the natural exponential family, $\EF(\eta; b, \psi)$, with density
\begin{equation}\label{eq:EF}
	p(Y \given \eta) = \exp \{ \eta Y - b \psi(\eta) + c(Y) \}, \ Y \in \mathcal{Y}, \eta \in \mathcal{H}\;,
\end{equation}
where $\mathcal{Y}$ denotes the support of $Y$ and $\mathcal{H} = \{\eta : \psi(\eta) < \infty \}$ denotes the natural parameter space and, therefore, the support of $\eta$. The scalar $b$ may be unknown, while $\psi(\cdot)$ and $c(\cdot)$ are known functions. We will discuss various forms of the unit log partition function $\psi(\cdot)$ and the scalar $b$ to denote different distributions of the family. For example, $\psi_1(t) = t^2$ with $b = 1/2\sigma^2$, $\psi_2(t) = e^t$ with $b = 1$ and $\psi_3(t) = \log (1 + e^t)$ with $b = 1$ or $m > 1$ correspond to the Gaussian with variance $\sigma^2$, Poisson, binary or binomial distributions, respectively. 

\cite{DY79} proposes a proper conjugate prior for $\eta$ in \eqref{eq:EF}, 
\begin{equation}\label{eq:DYdef}
	p(\eta \given \alpha, \kappa) \propto \exp \{ \alpha \eta - \kappa \psi(\eta) \}, \ \eta \in \, \frac{\alpha}{\kappa} \in \mathcal{Y}, \kappa > 0 \;,
\end{equation}
which we denote as $\eta \sim \DY(\alpha, \kappa; \psi)$. It follows that $\eta \given Y, \alpha, \kappa \sim \DY(\alpha + Y, \kappa + b; \psi)$. Special cases of the $\DY$ distribution include Gaussian ($\psi = \psi_1$), log-gamma ($\psi = \psi_2$), and logit-beta ($\psi = \psi_3$). Some may not belong to the exponential family. For example, $\alpha = 0$, $\psi(t) = \psi_4(t, \nu) = \log(1 + t^2/\nu)$, and $\kappa = (\nu + 1)/2$ with $\nu > 0$ yield a Student's $t$-distribution with $\nu$ degrees of freedom.

\subsection{The Generalised Conjugate Multivariate distribution}\label{subsec:CMdist}
We modify the generalised conjugate multivariate random variable ($\GCM$) introduced by \cite{bradley2023lgp} using linear combinations of mutually independent DY random variables. We define $\zeta$ as the $n\times 1$ random vector
\begin{equation}\label{eq:CMtrans}
	\zeta = \mu + L D(\theta) \eta \;,
\end{equation}
where the location parameter $\mu$ is $n\times 1$, $\eta = (\eta_1, \dots, \eta_n)^\T$ with elements $\eta_i \overset{\text{ind}}{\sim} \DY(\alpha_i, \kappa_i; \psi_i)$ with $\kappa_i > 0$ for $i=1,\ldots,n$, $L$ is $n \times n$ lower-triangular with positive diagonal elements, $\theta$ denotes a set of scale parameters with parameter space $\Theta$, and $D(\theta)$ is an $n \times n$ invertible matrix for all values of $\theta \in \Theta$. Given $\theta$, we use \eqref{eq:DYdef} to derive the conditional density of $\zeta$ up to a proportionality constant,
\begin{equation}\label{eq:CMdens}
\begin{split}
    p(\zeta \given \mu, L, \theta, \alpha, \kappa; \psi) &\propto \exp\left\{\alpha^\T D(\theta)^{-1} L^{-1} (\zeta - \mu) \right. \\
    & \left. \quad - \kappa^\T \psi ( D(\theta)^{-1} L^{-1} (\zeta - \mu) ) \right\} \det(L^{-1}) 
    \det(D(\theta)^{-1}) \;,
\end{split}
\end{equation}
where $\psi = (\psi_1, \ldots, \psi_n)^{\T}$, and $\psi(D(\theta)^{-1}L^{-1}(\zeta - \mu))$ is an $n \times 1$ vector with its $i$th element $\psi_i(e_i^\T D(\theta)^{-1} L^{-1}(\zeta - \mu))$, $e_i$ is the vector of zeros with 1 at the $i$th position, $\alpha = (\alpha_1, \dots, \alpha_n)^\T$ and $\kappa = (\kappa_1, \dots, \kappa_n)^\T$ are $n\times 1$ shape and scale parameters. The $\GCM$ is derived from \eqref{eq:CMdens} considering a probability law for $\theta$ independent of the remaining parameters, i.e., $\pi(\theta) := \pi(\theta\given \mu, L, \alpha, \kappa)$. We write $\zeta \sim \GCM(\mu, L, \alpha, \kappa, D, \pi; \psi)$ to denote $p(\zeta \given \mu, L, \alpha, \kappa; \psi) \propto \E_{\theta}\left[p(\zeta \given \mu, L, \theta, \alpha, \kappa; \psi)\right]$. 

Moreover, suppose $\zeta = (\zeta_1^\T, \zeta_2^\T)^\T$, where $\zeta_1$ is $r\times 1$ and $\zeta_2$ is $(n-r)\times 1$. Then, given $\theta$, the conditional density of $\zeta_1$ given $\zeta_2$ up to a proportionality constant is
\begin{equation}\label{eq:cCMdens}
\begin{split}
p(\zeta_1 \given \zeta_2 = c_2, \mu^*, A_1, \theta, \alpha, \kappa; \psi) \propto & \exp\{ \alpha^\T (D(\theta)^{-1} A_1 \zeta_1 - \mu^*) \\ 
& \quad - \kappa^\T \psi (D(\theta)^{-1} A_1 \zeta_1 - \mu^*) \} \det(D(\theta)^{-1}) \;,
\end{split}
\end{equation}
where $A_1$ is the $n \times r$ submatrix of $L^{-1} = [A_1, A_2]$, and $\mu^* = D(\theta)^{-1} L^{-1}\mu - D(\theta)^{-1} A_2 c_2$ for some $c_2 \in \mathbb{R}^{n-r}$. The proportionality constant in \eqref{eq:cCMdens} is strictly positive and finite, ensuring that \eqref{eq:cCMdens} is proper \citep{bradleyholanwikle}. Under the same assumptions of $\pi(\theta)$ as above, we write $\zeta_1 \given \zeta_2 \sim \GCMc(\mu^*, A_1, \alpha, \kappa, D, \pi; \psi)$ for $p(\zeta_1 \given \zeta_2, \mu, L, \alpha, \kappa; \psi) \propto {\E_{\theta \given \zeta_2, \mu, L, \alpha, \kappa; \psi}\left[p(\zeta_1 \given \zeta_2, \mu, L, \theta, \alpha, \kappa; \psi)\right]}$.

We use these distributions to build hierarchical models in the following sections. In general, we may be unable to sample directly from the conditional $\GCM$ distribution except for some familiar exceptions (e.g., the conditional distribution of Gaussian is indeed Gaussian). However, it is possible to consider an augmented model with a particular structure that yields a posterior distribution in the $\GCM$ family that is easy to sample from (Section~\ref{sec:bayeshier}).

\section{Bayesian hierarchical model}\label{sec:bayeshier}
\subsection{Conjugate Bayesian spatially-temporally varying coefficients model}\label{subsec:conjugate}
Let $\calL = \{\ell_1, \allowbreak \ldots, \ell_n\}$ be a fixed set of $n$ distinct space-time coordinates in $\calD$, where $y(\calL) = (y(\ell_1), \dots, y(\ell_n))^\T \in \mathcal{Y}^n$, which we simply denote by $y$, is the vector of observed outcomes, each distributed as a member of the natural exponential family with log partition function $\psi_y$. Suppose, $x(\ell_i)$ is a $p\times 1$ vector of predictors, $\beta$ is the corresponding $p \times 1$ vector of slopes (fixed effects), $\tilde{x}(\ell_i)$ is $r\times 1$ ($r \leq p$) consisting of predictors in $x(\ell_i)$ that are posited to have spatially-temporally varying regression coefficients $z(\ell_i) = (z_1(\ell_i), \ldots, z_r(\ell_i))^\T$, where each $z_j(\ell_i)$ is a spatially-temporally varying coefficient for the predictor $\Tilde{x}_j(\ell_i)$, $\xi_i$ is a fine-scale variation term and $\mu_i$ is the discrepancy parameter \citep{bradleyholanwikle}. We introduce spatially-temporally varying coefficients in $\eta(\ell)$ as 
\begin{equation}\label{eq:model_final}
\begin{split}
    y(\ell_i) &\given \beta, z(\ell_i), \xi_i, \mu_i \overset{\text{ind}}{\sim} \EF \left(\eta(\ell_i) + \xi_i - \mu_i; b_i, \psi_y \right), \ i=1,\ldots,n\;,\\
    \eta(\ell) &= x(\ell)^\T \beta + \tilde{x}(\ell)^{\T}z(\ell), \quad  \beta \given \sigma^2_\beta, \mu_\beta, V_\beta \sim \Norm (\mu_\beta, \sigma^2_\beta V_\beta), \quad \sigma^2_\beta \sim \pi_\beta(\sigma^2_\beta) \;,\\
    z(\ell) &\given \theta_z, \thetasp \sim \GP (0, C_z(\cdot, \cdot; \thetasp, \theta_z))\;,\quad \theta_z \sim \pi_{z}(\theta_z)\;, \\
    \xi &\given \beta, z, \mu, \alpha_\epsilon, \kappa_\epsilon, \sigma^2_\xi \sim \GCMc (\tilde{\mu}_\xi, H_\xi, \alpha_\xi, \kappa_\xi, D_\xi, \pi_\xi; \psi_\xi), \ \sigma^2_\xi \sim \pi_\xi(\sigma^2_\xi), \ p(\mu) \propto 1 \;,
\end{split}
\end{equation}
where $z(\ell) = (z_1(\ell), \ldots, z_r(\ell))^\T$ is a multivariate Gaussian process with a separable cross-covariance function $C_z(\cdot, \cdot; \thetasp, \theta_z)$, characterised by the process parameters $\thetasp$ that control the spatial-temporal correlation within the process and $\theta_z$ that control the covariance matrix between the processes \citep{mardiagoodall1993}. Given $\theta_z$, the $nr \times 1$ vector $z = (z_1^\T, \ldots, z_r^\T)^\T$, where $z_j = (z_j(\ell_1), \ldots, z_j(\ell_n))^\T$ for $j = 1, \ldots, r$, follows a multivariate Gaussian distribution with mean $0$ and $nr \times nr$ covariance matrix $C_z(\calL; \thetasp, \theta_z)$. We discuss different specifications for $C_z$ in Section~\ref{subsec:spt_process}.

Let  $X$ be $n \times p$ with the value of the $j$th predictor at $\ell_i$ as the $(i,j)$-th element $x_j(\ell_i)$ for $j=1,\ldots,p$, and $\Tilde{X} = [\mathrm{diag}(\Tilde{x}_1), \ldots , \mathrm{diag}(\Tilde{x}_r)]$ is $n \times nr$ with $\mathrm{diag}(\Tilde{x}_j)$ being an $n \times n$ diagonal matrix whose $i$th diagonal element is $\Tilde{x}_j(\ell_i)$ for $j = 1, \ldots, r$. The conditional prior for $\xi$ is characterised by the $2n \times 1$ location $\tilde{\mu}_\xi = ((\mu - X \beta - \Tilde{X} z)^\T, 0_n^\T )^\T$, where $0_k$ denotes the vector of zeros of length $k \in \mathbb{N}$, $2n \times n$ matrix $H_\xi = [I_n, I_n]^\T$, $2n \times 2n$ block-diagonal matrix $D_\xi(\sigma^2_\xi) = \mathrm{blkdiag}(I_n, \sigma_\xi I_n)$. The shape parameter $\alpha_\xi = (\alpha_\epsilon 1_n^\T, 0_n^\T)^\T$ and the scale parameter $\kappa_\xi = (\kappa_\epsilon 1_n^\T, (1/2)1_n^\T)^\T$, where $1_k$ is the vector of ones of length $k \in \mathbb{N}$. Finally, we assume $p(\mu) \propto 1$. 

Instead of applying MCMC to sample from \eqref{eq:model_final}, we devise predictive stacking that averages over models specified by fixing parameters and hyperparameters that serve to tune inference but are not, by themselves, of interest. Therefore, let $M$ denote a generic version of \eqref{eq:model_final} obtained by specifying fixed values for $\thetasp$, the boundary adjustment parameters $\{\alpha_\epsilon, \kappa_\epsilon\}$, hyperparameters $\mu_\beta$, $V_\beta$ and other auxiliary hyperparameters present in the hyperprior $\pi(\theta) = \pi_z(\theta_z) \, \pi_\beta(\sigma^2_\beta) \, \pi(\sigma^2_\xi)$ for $\theta = \{\theta_z, \sigma^2_\beta, \sigma^2_\xi\}$. These specifications ensure analytically accessible posterior distributions from the $\GCM$ family that are easy to sample from. The model in \eqref{eq:model_final} collapses into a parsimonious representation by integrating out the scale parameters $\theta$ with respect to the prior $\pi(\theta)$. This yields
\begin{equation}\label{eq:model_concise}
\begin{split}
    y(\ell_i) \given \beta, z(\ell_i), \xi_i, \mu_i &\overset{\text{ind}}{\sim} \EF \left(\eta(\ell_i) + \xi_i - \mu_i; b_i, \psi_y \right), \ i = 1, \dots, n \\
    (\gamma^\T, q^\T)^{\T} \given M &\sim \GCM (0_{2n+p+nr}, V, \alpha, \kappa, D, \pi; \psi) \;,
\end{split}
\end{equation}
where $\gamma = (\xi^\T, \beta^\T, z^\T)^\T$, with $n \times 1$ vector $\xi = (\xi_1, \ldots, \xi_n)^\T$ and $q$ is a reparametrization of $\mu$, which we detail below. The matrix $V$ is $(2n+p+nr)\times (2n+p+nr)$ such that $V^{-1} = [H, Q]$, where $H = \begin{bmatrix}
    I_n & X& \Tilde{X}\\ I_n & 0 & 0 \\ 0 & L_\beta^{-1} & 0 \\ 0 & 0 & L_z^{-1}
\end{bmatrix}$ 
is $(2n+p+nr)\times (n+p+nr)$ and $Q$ is $(2n+p+nr)\times n$ whose columns are the $n$ orthonormal eigenvectors of $I - P_H$ corresponding to eigenvalue $1$ with $P_H = H (H^\T H)^{-1} H^\T$ being the orthogonal projection matrix on the column space of $H$,   
$L_\beta$ is the lower-triangular Cholesky factor of $V_{\beta}$ and $L_z$ is an $nr\times nr$ lower-triangular matrix obtained from our specification of $C_z({\calL}; \thetasp, \theta_z)$ (see Section~\ref{subsec:spt_process} for details on $L_z$). 

We define $q = - Q^\T D(\theta) \Tilde{\mu}$, where $\tilde{\mu} = (\mu^\T, 0_n^\T, \mu_\beta^\T L_\beta^{-\T}, 0_{nr}^\T)^\T$ and $\mu = (\mu_1, \ldots, \mu_n)^\T$ . Here, $D(\theta) = \mathrm{blkdiag}(I_n, \sigma_\xi I_n, \sigma_\beta I_p, D_z(\theta_z))$. Finally, the shape and scale parameters of $\GCM$ in \eqref{eq:model_concise} are $\alpha = (\alpha_\epsilon 1_n^\T, 0_{n+p+nr}^\T)^\T$, and, $\kappa = (\kappa_\epsilon 1_n^\T, (1/2) 1_{n+p+nr}^\T)^\T$. The unit log partition function $\psi(\cdot)$ in \eqref{eq:model_concise} is $\psi (h) = (\psi_y(h_1)^{\T}, \psi_1(h_2)^{\T})^{\T}$ for some $h = (h_1^{\T}, h_2^{\T})^{\T}$ with $h_1 \in \mathbb{R}^n$, $h_2 \in \mathbb{R}^{n + p + nr}$, where all log partition functions operate element-wise on the arguments. See Lemma~\ref{lemma:rank_supp} and Theorem~\ref{thm:improper_prior_supp} in Appendix~\ref{sec:posterior_supp} for technical details.

\subsection{Posterior distribution}
The hierarchical model \eqref{eq:model_final} (or \eqref{eq:model_concise}) yields the posterior distribution
\begin{equation}\label{eq:posterior_final}
    (\gamma^\T, q^\T)^\T \given y, M \sim \GCM (0_{2n+p+nr}, V, \alpha^*, \kappa^*, D, \pi; \psi ) \;,
\end{equation}
where the shape and scale parameters are $\alpha_\ast = ((y + \alpha_\epsilon 1_n)^\T, 0_{n+p+nr}^\T)^\T$ and $\kappa_\ast = ((b + \kappa_\epsilon 1_n)^\T, \allowbreak (1/2) 1_{n+p+nr}^\T)^\T$ (Theorem~\ref{thm:posterior_supp} in Appendix~\ref{sec:posterior_supp}). Generalised linear models typically assume $\mu$ (and hence $q$) to be zero. This yields $p(\gamma \given y, M)$ to be a $\GCMc$ distribution, which we cannot sample directly from, except for some special cases of $\psi_y$ (e.g., Gaussian). We note that $\mu$ is crucial in producing the posterior distribution within the $\GCM$ family and therefore, unlike traditional generalised linear models, cannot be excluded from \eqref{eq:model_final}. Parameters $\{\alpha_\epsilon, \kappa_\epsilon\}$ ensure that the posterior shape and scale parameters $\alpha^\ast$ and $\kappa^\ast$ do not lie on the boundary of the parameter space, and hence the inclusion of $\xi$ is necessary for a well-defined posterior distribution.

\subsection{Fixed effects and spatial-temporal processes}\label{subsec:spt_process}
In practice, we have found $\sigma^2_\xi$ to have little impact on posterior inference and therefore fix it to a small positive number, i.e., $\pi(\sigma^2_\xi) = \delta_{c}$ for some $c > 0$, where $\delta_c$ corresponds to the Dirac measure on a real $c$. We also assign an inverse-gamma prior $\pi(\sigma_\beta^2) = \IG( \sigma^2_\beta \given \nu_\beta/2, \nu_\beta/2)$. 
Similarly, we model $z$ using a spatial-temporal Gaussian process, and subsequently place the inverse-gamma/inverse-Wishart prior on the scale parameter $\theta_z$. Below, we discuss some possible spatial-temporal process models to study the spatially-temporally varying effects of the predictors on the response.

\subsubsection{Independent spatial-temporal processes}
We consider $r$ independent Gaussian spatial-temporal processes
\begin{equation}\label{eq:z_ind}
\begin{split}
    \text{Independent processes: } z_j(\ell) \given \sigma_{z_j}^2, \thetasp_j & \overset{\text{ind}}{\sim} \GP (0, \sigma_{z_j}^2 R_j(\cdot, \cdot; \thetasp_j)),\\
    \sigma_{z_j}^2 & \sim \IG(\nu_{z_j}/2, \nu_{z_j}/2), \quad j = 1, \ldots, r,
\end{split}
\end{equation}
where $\sigma_{z_j}^2$ is the variance parameter corresponding to the process $z_j(\ell)$. This corresponds to the covariance matrix $C_z(\calL; \thetasp, \theta_z) = \oplus_{j = 1}^r \sigma_{z_j}^2 R_j(\thetasp_j)$ in \eqref{eq:model_final} with $\thetasp = \{ \thetasp_j : j = 1, \ldots, r\}$, where $\thetasp_j$ denotes the covariance kernel parameters for the $j$th process and $\theta_z = \{\sigma^2_{z_1}, \ldots, \sigma^2_{z_r}\}$. We let $L_z$ be the $nr\times nr$ lower-triangular Cholesky factor of the block diagonal matrix $\oplus_{j = 1}^r R_j(\thetasp_j)$ and $D_z(\theta_z) = \oplus_{j = 1}^r \sigma_{z_j} I_n$, such that $C_z(\calL; \thetasp, \theta_z) = L_z D_z D_z^{\T} L_z^{\T}$. 
Lastly, \eqref{eq:z_ind} places independent inverse-gamma priors on each scale parameter, given by $\pi(\theta_z) = \prod_{j = 1}^r \IG(\sigma^2_{z_j} \given \nu_{z_j}/2, \nu_{z_j}/2)$.

\subsubsection{Multivariate spatial-temporal process}
We can introduce dependence among the elements of the $r\times 1$ vector $z(\ell)$ using
\begin{equation}\label{eq:multi_z}
    \text{Multivariate process: }z(\ell) \given \Sigma \sim \GP (0, R(\cdot, \cdot; \thetasp)\Sigma), \quad \Sigma \sim \IW(\nu_z + 2r, \Psi)\;,
\end{equation}
where $\mathcal{GP} (0, R(\cdot, \cdot; \thetasp)\Sigma)$ is an $r\times 1$ multivariate Gaussian process with matrix-valued cross-covariance function $R(\cdot, \cdot; \thetasp)\Sigma$ and $\Sigma$ is an $r \times r$ positive definite random matrix. This corresponds to the spatial-temporal covariance matrix $C_z(\calL; \thetasp, \theta_z) = \Sigma \otimes R(\thetasp)$ in \eqref{eq:model_final} with $\theta_z = \Sigma$. Here, we let $L_z$ be the $nr \times nr$ lower-triangular Cholesky factor of the block-diagonal matrix $I_r \otimes R(\thetasp)$ and $D_z(\theta_z) = \Sigma^{1/2} \otimes I_n$, where $\otimes$ denotes the Kronecker product. Hence, given $\Sigma$, the $n \times r$ matrix $Z = [z_1 : \ldots : z_r]$ follows a matrix normal distribution $\MNorm_{n, r}(0, R(\thetasp), \Sigma)$ \citep[][Chapter~2.2]{guptanagar_matrixvariate}. We place an inverse-Wishart prior on the scale parameter with the shape $\nu_z + 2r$ and $r\times r$ positive definite scale matrix $\Psi$ \citep[see, e.g.,][Chapter~3.4]{guptanagar_matrixvariate}, given by $\pi(\theta_z) = \IW (\Sigma \given \nu_z + 2r, \Psi)$.

\subsubsection{Spatial-temporal correlation kernel}
For independent processes, elements of $R_j(\thetasp_j)$ are evaluated using a spatial-temporal correlation function \citep[see, e.g.,][]{gnei10} 
\begin{equation}\label{eq:corr_fn}
    R_j(\ell, \ell'; \thetasp_j) = \frac{1}{ \phi_{1j} |t-t'|^2+1 } \exp\left(  -\frac{\phi_{2j} \| s-s' \| }{\sqrt{1+\phi_{1j} |t-t'|^2} }   \right),\quad \phi_{1j},\phi_{2j}>0 \;,
\end{equation}
where $\ell=(s,t)$ and $\ell'=(s',t')$ are any two distinct space-time coordinates in $\calD$, $\|\cdot\|$ is the Euclidean distance over $\calS$, $\thetasp_j = (\phi_{1j}, \phi_{2j})$, and $\phi_{1j}$ and $\phi_{2j}$ are positive spatial and temporal decay parameters, respectively. We gather the $2r$ process parameters in $\thetasp = \{\thetasp_j: j = 1, \ldots, r\}$. For ease of notation, we drop $\thetasp_j$ and simply write $R_j(\cdot, \cdot)$ for the correlation function. {F}or the multivariate model in \eqref{eq:multi_z}, $R(\cdot, \cdot; \thetasp)$ corresponds to \eqref{eq:corr_fn} with $\thetasp = \{\phi_1, \phi_2\}$.

\section{Predictive Stacking}\label{sec:stacking}

\subsection{Choice of candidate models}\label{subsec:candidate}
We consider modelling data typically originating from Poisson ($\psi_y = \psi_2$), binomial and binary ($\psi_y = \psi_3$) distributions. In order to implement stacking, we first fix values of certain hyperparameters. In practice, we assume $\mu_\beta = 0$, $V_\beta = \delta_\beta^2 I_p$ for some real $\delta_\beta$, and some $\nu_\beta > 2$ to specify the priors of fixed effects $\beta$ and $\sigma_\beta^2$. Moreover, under the assumption of independent processes, we assume $\nu_{z_j} > 2$ for each $j$, and, on the other hand, $\nu_z > 2$ and some positive definite $\Psi$ for the assumption of a multivariate process on varying coefficients.

Once the hyperparameters are fixed, we consider a grid of candidate values for the spatial process parameters $\thetasp$ and the boundary adjustment parameter $\alpha_\epsilon$, conditional on which we sample from the posterior \eqref{eq:posterior_final}. We choose $\alpha_\epsilon > 0$ in the prior of $\xi$ to ensure that $\alpha^*$ in \eqref{eq:posterior_final} does not lie on the boundary of its parameter space. For Poisson data, $\kappa_\epsilon = 0$ and for binary/binomial data, $\kappa_\epsilon = 2 \alpha_\epsilon$. A natural approach to construct a collection of candidate models $\calM$ is to consider a Cartesian product of candidate values for each parameter in $\{\thetasp, \alpha_\epsilon\}$. Although that is computationally viable under the multivariate specification \eqref{eq:multi_z}, where $\thetasp = \{\phi_1, \phi_2\}$ is of a lower dimension compared to that of independent processes \eqref{eq:z_ind}, where $\thetasp = \{(\phi_{1j}, \phi_{2j}): j = 1, \ldots, r\}$ is $2r$-dimensional. In the latter case, the number of models, when naively constructed, grows exponentially, undermining the computational advantage gained by sampling from analytically available posterior distributions. Instead, in this case, we build $\calM$ using a fixed number of random candidate values of $\thetasp$, sampled uniformly from a $2r$-dimensional rectangle specified by plausible lower and upper bounds for each parameter, likely supplied by the user.

\subsection{Sampling from posterior}\label{subsec:sampling}
Sampling from the posterior distribution \eqref{eq:posterior_final} proceeds by first sampling replicates from the family of marginal priors for $\gamma$ obtained by integrating out the scale parameters. We elaborate in the following. Since $[H, Q]^{-1} = [H(H^\T H)^{-1}, Q]^\T$, we compute
\begin{equation}\label{eq:proj1}
    \gamma^{(b)} = (H^\T H)^{-1} H^\T v^{(b)} \quad \mbox{and}\quad q^{(b)} = Q^\T v^{(b)}
\end{equation}
to obtain $(\gamma^{(b) \T}, {q}^{(b) \T})^\T$, $b = 1, \ldots, B$ of $(\gamma^\T, q^\T)^\T$, where $v^{(b)}$ is a sample from the distribution $\GCM (0, I_{2n+p+nr}, \alpha^*, \kappa^*, D, \pi; \psi_\gamma)$. The random vector $v^{(b)} = (v_\eta^{(b)\T}, v_\xi^{(b)\T}, v_\beta^{(b)\T}, \allowbreak v_z^{(b)\T})^\T$ is made up of $v_\eta^{(b)}$, where $v^{(b)}_{\eta, i} \sim \DY(y_i + \alpha_\epsilon, b+\kappa_\epsilon; \psi_y)$ for $i = 1, \ldots, n$ and $v^{(b)}_\beta$, $v_z^{(b)} = (v_{z_1}^{(b)\T}, \ldots, v_{z_r}^{(b)\T})^{\T}$, and $v^{(b)}_\xi$ is a sample from $\mathrm{N}(0, \sigma_\xi^2 I_n)$ for each $b$. The $n\times 1$ random vector $v^{(b)}_\eta$ has its $i$-th element as $\eta^{(b)}(\ell_i) + \xi^{(b)}_i - \mu^{(b)}_i$. 

The projection in \eqref{eq:proj1} maps the posterior samples of $\eta$ to the posterior samples of $\gamma$ accounting for the effect of its priors. It is instructive to rewrite \eqref{eq:proj1} as 
\begin{equation}\label{eq:proj2}
    \gamma^{(b)} = ( H_1^{\T} H_1 + V_\gamma^{-1} )^{-1} ( H_1^\T v^{(b)}_\eta + L_\gamma^{-\T} v^{(b)}_{\gamma} )\;,
\end{equation}
where $H = [H_1^\T, L_\gamma^{-\T}]^\T$ with $H_1 = [I_n, X, \Tilde{X}]$, $V_\gamma = L_\gamma L_\gamma^\T$, and $v^{(b)}_{\gamma} = (v_\xi^{(b)\T}, v_\beta^{(b)\T}, v_z^{(b)\T})^\T$. We leverage the structure of $H$ and the sparsity of $\Tilde{X}$ to achieve a computationally efficient algorithm for evaluating \eqref{eq:proj2}. Section~\ref{sec:projalgo} of the Appendix summarises an optimised algorithm for computing \eqref{eq:proj2} given a $v^{(b)}$. Moreover, the posterior samples of the scale parameters are recovered from the posterior samples of $\{\beta^{(b)}, z^{(b)}\}$ from the marginal model (Section~\ref{subsec:recovery} in the Appendix).

Posterior predictive inference for the spatial-temporal process is carried over to the canonical mean $\eta(\ell) = x(\ell)^\T \beta + \Tilde{x}(\ell)^{\T} z(\ell)$. For each value of $\{\beta^{(b)}, z^{(b)}(\ell)\}$ drawn from the posterior distribution \eqref{eq:posterior_final}, we obtain the corresponding sample $\eta^{(b)}(\ell) = x(\ell)^\T \beta^{(b)} + \Tilde{x}(\ell)^{\T} z^{(b)}(\ell)$. This can be achieved at any arbitrary space-time coordinate $\ell$. If $\ell \in \calL$, then we can further generate model replicated data $Y_{\text{rep}}^{(b)}(\ell_i) \sim \EF (\eta^{(b)}(\ell_i) + \xi_i^{(b)} - \mu_i^{(b)}; b_i, \psi_y )$ for each posterior draw of $\{\eta^{(b)}(\ell_i), \xi_i^{(b)}, \mu_i^{(b)}\}$. If, on the other hand, $\ell \notin \calL$, then we seek predictive inference. Here, we can still carry out inference for $\eta(\ell)$ but cannot coherently extend inference to the outcome $Y(\ell)$. This is an artefact of spatial GLMs, where the well-defined process $z(\ell)$ in $\eta(\ell)$ does not extend to the outcome. Although we have a well-defined joint distribution for outcomes on a finite set of points, we do not have a corresponding stochastic process $Y(\ell)$ across the domain \citep[][]{banerjee_spatial}. Section~\ref{subsec:prediction} offers posterior predictive inference by defining predictive random variables $\tilde{Y}(\Tilde{\ell}) \sim \EF \left(\eta(\Tilde{\ell}); \tilde{b}, \psi_y \right)$. 

\subsection{Spatial-temporal prediction}\label{subsec:prediction}
Given the data observed at $\calL$, let $\Tilde{\calL} = \{ \Tilde{\ell}_1, \dots, \Tilde{\ell}_{\Tilde{n}} \} \subset \mathcal{D} \setminus \calL$ be a collection of $\Tilde{n}$ new space-time coordinates in $\mathcal{D}$, where we wish to predict the response and the latent spatial-temporal processes. Let $\Tilde{y}$ and $\Tilde{z}_j$ for each $j$ be the $\Tilde{n} \times 1$ vectors with the $i$th elements $y(\Tilde{\ell}_i)$ and $z_j(\Tilde{\ell}_i)$, respectively. For a given model, $M$, which entails a fixed value of $\thetasp$ and some auxiliary model parameters, spatial-temporal predictive inference evaluates the posterior predictive distribution,
\begin{equation}\label{eq:pred}
    p(\Tilde{y}, \Tilde{z} \given y, M) = \int p(\Tilde{y} \given \beta, \Tilde{z}) p(\Tilde{z} \given z, M) p(\beta, z \given y, M) d\beta dz\;,
\end{equation}
where $\Tilde{z} = (\Tilde{z}_1^{\T}, \ldots, \Tilde{z}_r^{\T})^{\T}$. We sample from \eqref{eq:pred} first drawing $\{ \beta^{(b)}, z^{(b)} \}$ from $p(\beta, z \given y, M)$ as described in Section~\ref{subsec:sampling} and then, for each drawn value $z^{(b)}$ from $z$, drawing $\Tilde{z}^{(b)}$ from $p(\Tilde{z} \given z, M)$. Furthermore, for each posterior sample $\beta^{(b)}$ and $\Tilde{z}^{(b)}$, we draw $\Tilde{y}^{(b)}$ from $p(\Tilde{y} \given \beta, \Tilde{z})$ with $\mu$ in \eqref{eq:model_final} set to 0. This yields $\{\Tilde{y}^{(b)}, \Tilde{z}^{(b)}\}$ from \eqref{eq:pred}. 

For the independent process model in \eqref{eq:z_ind}, the marginal distribution of the $(n+\Tilde{n}) \times 1$ vector $(z_j^\T, \Tilde{z}_j^\T)^\T$ corresponding to the $j$th process is a multivariate $t$-distribution $t_{n+\Tilde{n}}(\nu_{z_j}, 0_n, \Tilde{V}_{z_j})$ with $\nu_{z_j}$ degrees of freedom, location parameter $0_n$ and $(n + \Tilde{n}) \times (n + \Tilde{n})$ scale matrix
$\displaystyle \Tilde{V}_z = \begin{bmatrix} R_j & C_j \\ C_j^{\T} & \Tilde{R}_j\end{bmatrix}$, where 
$\Tilde{R}_j$ is the $\Tilde{n}\times \Tilde{n}$ correlation matrix for $\tilde{z}_j$, and $C_j = [R_j(\ell, \ell')]$ is $n \times \Tilde{n}$ with $\ell \in \calL, \ell' \in \Tilde{\calL}$. This yields
\begin{equation}\label{eq:cond_t}
    \Tilde{z}_j \given z_j, M \sim t_{\Tilde{n}} \left(\nu_{z_j} + n, \, C_j^\T R_j^{-1} z_j, \, \frac{\nu_{z_j} + z_j^\T R_j^{-1} z_j}{\nu_{z_j} + n} ( \Tilde{R}_j - C_j^\T R_j^{-1} C_j ) \right),
\end{equation}
for $j = 1, \ldots, r$. It should be noted that the conditional scale matrix contains the factor $(\nu_{z_j} + z_j^\T R_j^{-1} z_j)/(\nu_{z_j} + n)$ which is directly related to the Mahalanobis distance of $z_j$, implying that the dispersion is enlarged in the presence of extreme values of $z_j$. The degrees of freedom also increase by a factor $n$, which means that the more data we have, the less heavy-tailed $p(\Tilde{z}_j \given z_j, M)$ becomes \citep{ding2016conditional}. 

For the multivariate process in \eqref{eq:multi_z}, the marginal distribution of the $(n+\Tilde{n}) \times r$ random matrix $[Z^\T, \Tilde{Z}^\T]^\T$, where $\Tilde{Z} = [\tilde{z}_1, \ldots, \tilde{z}_r]$ follows a matrix $t$-distribution $\matrixT_{n+\Tilde{n}, r}(\nu_{z}, 0,  \Tilde{V}_{z}, \Psi)$ with $(n + \Tilde{n}) \times (n + \Tilde{n})$ matrix 
$\displaystyle \Tilde{V}_z = \begin{bmatrix} R & C \\ C^{\T} & \Tilde{R}\end{bmatrix}$
defined accordingly ($R$, $C$ and $\Tilde{R}$ are common across $j$). We derive $p(\Tilde{Z} \given Z, M)$ as
\begin{equation}\label{eq:cond_MT}
    \Tilde{Z} \given Z, M \sim \matrixT_{\tilde{n}, r} \left(\nu_z + n, C^\T R^{-1} Z, \Tilde{R} - C^\T R^{-1} C, \Psi + Z^\T R^{-1} Z\right).
\end{equation}
For technical details, see Theorem~\ref{subsec:mv_spt} in the Appendix. Thus, the analytic tractability as described above provides further motivation behind the specifications of the spatial-temporal processes for practical purposes, since, an efficient way for evaluating the predictive density at an out-of-sample point is particularly crucial in order to effectuate our stacking algorithm.

\subsection{Stacking algorithm}\label{sec:stacking_algo}
We collect samples from $p(\beta, z \given \chi, M)$, where $\chi = \{y, X, \Tilde{X}, \calL\}$ denotes the data and $M$ specifies other parameters that must be conditioned for accessible posterior sampling. Following \citet{yao2018using}, we devise a stacking algorithm for \eqref{eq:model_final} based on predictive densities. Given a collection of candidate models $\mathcal{M} = \{M_1, \dots, M_G\}$, we find the probability distribution in $\mathcal{C} = \left\{\sum_{g = 1}^G w_g p(\cdot \given \chi, M_g) : \sum_{g = 1}^G w_g = 1, w_g \geq 0\right\}$ by solving for the optimal stacking weights $w = (w_1, \ldots, w_G)$ as the solution to the optimisation problem
\begin{equation}\label{eq:stack_optim}
    \max_{w_1, \dots, w_G} \frac{1}{n} \sum_{i = 1}^n \log \sum_{g = 1}^G w_g p(y(\ell_i) \given \chi_{-i}, M_g), \quad w_g \geq 0, \sum_{g = 1}^G w_g = 1 \;,
\end{equation}
where $\chi_{-i} = (y_{-i}, X_{-i}, \Tilde{X}_{-i}, \calL_{-i})$, i.e., the data excluding the $i$th observation.

Solving for $w$ in \eqref{eq:stack_optim} involves the leave-one-out predictive density $p(y(\ell_i) \given \chi_{-i}, M_g)$ for each $i$ and $g$. For each $g$, we require $n$ evaluations of the leave-one-out predictive densities to sample from \eqref{eq:posterior_final} under $M_g$, given data $\chi_{-i}$ for $i = 1, \dots, n$. Following \citet{vehtari_loo17}, we apply $K$-fold cross-validation for faster evaluation of these densities. We randomly permute the data and construct $K$ blocks using consecutive indices. Let $\chi_{[k]} = (y_{[k]}, X_{[k]}, \Tilde{X}_{[k]}, \calL_{[k]})$ be the $k$th block of size $n_k$ and $\chi_{[-k]} = \{y_{[-k]}, X_{[-k]}, \Tilde{X}_{[-k]}, \calL_{[-k]}\}$ be its complement of size $(n-n_k)$ for $k = 1,\dots,K$. For each $k$, we fit $M_g$ to $\chi_{[-k]}$ as we draw $S$ samples $\{\beta^{(s)}_{k,l}, z^{(s)}_{k,l}\}_{s = 1}^S \sim p((\gamma^\T, q^\T)^\T \given \chi_{[-k]}, M_g)$ as given in \eqref{eq:posterior_final}. 

Estimating $M_g$ on $\chi_{[-k]}$ requires a substantial computation dominated by Cholesky decompositions of matrices of size $\sim O(n)$ under both independent and multivariate process specifications. Under the independent process specification, instead of computing the Cholesky factor of $R_j(\calL_{[-k]})$ for every $\chi_{[-k]}$, which would cost $O(Kn^3)$ flops, we execute an efficient block Givens rotation \citep[][Section 5.1.8]{GolubLoanMatrix4} for faster evaluation of Cholesky factors of the $K$ submatrices of $R_j(\calL)$ taking $O(K n^3/4)$ operations for all blocks (see Algorithm~\ref{algo:cholesky} in the Appendix). Our algorithm is a block-level variant of \cite{kimcox2002_CV}. For each $z^{(s)}_{k, l} = (z^{(s) \T}_{1, k, l}, \ldots, z^{(s) \T}_{r, k, l})^{\T}$ where $z^{(s)}_{k, l}$ is $(n-n_k)r \times 1$, we draw posterior samples of $\Tilde{z}^{(s)}_{k, l} = (\Tilde{z}^{(s) \T}_{1, k, l}, \ldots, \Tilde{z}^{(s) \T}_{r, k, l})^{\T}$, the spatially-temporally varying $n_k r \times 1$ regression coefficients at the $n_k$ left-out locations, $\calL_{[k]}$ using \eqref{eq:cond_t} (see Section~\ref{subsec:prediction}), which we use to evaluate $p(y(\ell_i) \given \chi_{-i}, M_g)$. If $y(\ell_i) \in \chi_{[k]}$, then
\begin{equation}\label{eq:loo-pd}
    p(y(\ell_i) \given \chi_{-i}, M_g) \approx \frac{1}{S} \sum_{s = 1}^S \EF \left( y(\ell_i) \given x(\ell_i)^\T \beta^{(s)}_{k, l} + \Tilde{x}(\ell_i)^{\T} \Tilde{z}^{(s)}_{k, l}(\ell_i) ; b_i, \psi_y \right)\;,
\end{equation}
where $r \times 1$ vector $\Tilde{z}^{(s)}_{k, l}(\ell_i) = (\Tilde{z}^{(s)}_{1, k, l}(\ell_i), \ldots, \Tilde{z}^{(s)}_{r, k, l}(\ell_i))^{\T}$ and $\EF(y_0 \given \eta_0; b, \psi)$ is the density of $\EF(\eta_0; b, \psi)$ as defined in Section~\ref{subsec:dy} evaluated at $y_0 \in \mathcal{Y}$. We repeat these steps for each of the $G$ models and use \eqref{eq:loo-pd} to solve the optimisation problem \eqref{eq:stack_optim} using convex programming routines. Posterior inference for quantities of interest subsequently proceeds from the ``stacked posterior'',  
\begin{equation}
    \label{eq:stacked_posterior}
    \tilde{p}( \cdot \given \chi) = \sum_{g = 1}^G \hat{w}_g p(\cdot \given \chi, M_g)\;,
\end{equation}
where $\hat{w}_g$ are optimal weights from \eqref{eq:stack_optim}; see Algorithm~\ref{algo:stacking} in the Appendix.

Unlike MCMC, this stacking algorithm easily distributes independent tasks across multiple computing nodes and accrues substantial computational gains. The predictive stacking algorithm expends $O(G^{\alpha} + c^{-1} GKrn^3 / 4)$ flops for some $\alpha>0$, where $r$, $G$ and $K$ are as defined earlier and $c$ is the number of processing cores available in parallel. We use the package CVXR \citep{cvxr2020} in the \textsf{R} statistical computing environment by applying disciplined convex programming (2005 Stanford University Department of Electrical Engineering PhD thesis by M. Grant), while also confirming the convexity of the problem, and find the stacking weights in polynomial time $O(G^\alpha)$ using an interior-point algorithm. We use \textsf{Mosek} \citep{mosek} and \textsf{ECOSolveR} \citep{ecos} to calculate stacking weights.

\section{Simulation}\label{sec:simulation}
\subsection{Simulated data}\label{subsec:simulation}
We evaluate predictive performance of our proposed methods using two simulated datasets each of sample size $n$, with space-time coordinates sampled uniformly inside $[0, 1]^2 \times [0, 1]$, one with each varying coefficients are modelled as an independent spatial-temporal process, and the other assumes the varying coefficients are jointly a multivariate spatial-temporal process. We simulate the responses as Poisson count data $y(\ell) \sim \mathrm{Poisson}(\exp(\eta(\ell)))$, where $\eta(\ell) = x(\ell)^\T \beta + \tilde{x}(\ell)^\T z(\ell)$. In each case, $x(\ell)$ is $2\times 1$ consisting of an intercept and one predictor sampled from the standard normal distribution, and, the fixed effects $\beta = (5, -0.5)$. We set $\tilde{x}(\ell) = x(\ell)$, deeming all covariates to have spatially-temporally varying coefficients. Under the independent process assumption, $z_j(\ell) \sim \GP(0, \sigma^2_{z_j} R_j(\cdot, \cdot))$ for $j = 1, 2$ with $\sigma^2_{z_1} = 0.25$, $\sigma^2_{z_2} = 0.5$ and, $(\phi_{11}, \phi_{21}) = (0.5, 2)$ and, $(\phi_{12}, \phi_{22}) = (1, 4)$. For the multivariate case, we assume $\Sigma = [(2, 0.5)^\T, (0.5, 2)^\T]^\T$ and $\phi_1 = 1$ and $\phi_2 = 2$. The choice of $\beta$ in each simulated data is such that the generated data do not contain excessive zeros. The simulation experiments are conducted with $n$ varying from 200 to 600 with a randomly chosen holdout sample of size $n_h = 100$ over a set of coordinates $\calL_h$.

\subsection{Posterior inference}\label{subsec:inference}
We fit multivariate and independent process models to our data. In each case, we stack on the parameters $\{\thetasp, \alpha_\epsilon\}$, where $\thetasp$ denotes the collection of process parameters corresponding to spatial-temporal process model and $\alpha_\epsilon$ is the boundary adjustment parameter. For the prior on $\beta$, we consider $\mu_\beta = 0$ and, $V_\beta = I_2$. For subsequent inference, we fix hyperparameters $\nu_\beta = 2.1$ and $\nu_z = \nu_{z_1} = \nu_{z_2} = 2.1$. This ensures finite second moment of the corresponding marginal prior distributions of $\beta$ and $z$. The parameters $\alpha_\epsilon$ and $\kappa_\epsilon$ in \eqref{eq:model_final} specify the shape and scale parameters ($\alpha^*$ and $\kappa^*$) of the posterior distribution \eqref{eq:posterior_final}. 
If the data is on the boundary of the parameter space with a high frequency (e.g., Poisson data with excessive zeros), then inference can be sensitive to these parameters \citep{bradley2023lgp}. 

We formally handle the value of these parameters by stacking on several models with different choices of $\alpha_\epsilon$ (since $\kappa_\epsilon$ is uniquely determined by $\alpha_\epsilon$ for Poisson/binomial count and binary data). In each case, we fix the grid of boundary adjustment parameter as $G_{\alpha_\epsilon} = \{0.5, 0.75\}$. Finally, for fitting the multivariate, we choose candidate values of the spatial-temporal decay parameters so that the corresponding effective range is approximately between 20\% and 80\% of the maximum spatial and temporal inter-coordinate distances \citep[see][Chapter 2]{banerjee_spatial}, $G_{\phi_1} = \{0.3, 0.7, 1.2\}$ and $G_{\phi_2} = \{0.5, 0.75, 1.5\}$. In this case, we consider a Cartesian product of the grids, which yields $2 \times 3 \times 3 = 36$ candidate tuples of $\{\thetasp, \alpha_\epsilon\}$, and, hence, we stack on 36 models. For the simulated data with independent process specification, we instead sample 36 points $\{(\phi_{11}, \ldots, \phi_{1r}), (\phi_{21}, \ldots, \phi_{2r}), \alpha_\epsilon\}$ uniformly from $[0.3, 1.3]^r \times [0.4, 1.5]^r \times G_{\alpha_\epsilon}$. We use $K = 10$ for $K$-fold cross validation \citep{vehtariCV2002}.

We also estimate a fully Bayesian model with prior distributions on the spatial process parameters using MCMC for comparison with predictive stacking. In addition to the same priors for the model parameters, as mentioned above, we assign uniform priors $\mathrm{U}(0.1, 5)$ on all process parameters in $\thetasp$. Here, it is worth remarking that sampling from the joint posterior distribution, for example using random walk Metropolis steps, suffers considerably from mixing and convergence issues because of the high-dimensional parameter space and weak identifiability of process parameters. This issue is only partially mitigated using adaptive Metropolis steps \citep{adaMetropGibbs_2009}. The Gibbs sampling algorithm involves sampling from the conditional posterior distributions $p(\thetasp \given \beta, z, \xi, y)$ and $p(\beta, z, \xi, \given \thetasp, y)$, where the former involves an adaptive Metropolis update using the \textsf{R} package \textsf{spBayes} \citep{spBayes_r} and, the latter proceeds by sampling from \eqref{eq:posterior_final} using the projections described in Section~\ref{subsec:sampling}. Furthermore, we compare predictive stacking with the integrated nested Laplace approximation \citep[INLA;][]{inla2009}, a widely used method for approximate Bayesian inference. We implement our model in INLA under the conventional assumption that the discrepancy parameter $\mu$ equals zero. The computation is carried out within the \textsf{R} package \textsf{INLAspacetime} \citep{INLAspacetime_r}, which uses the \texttt{cgeneric} interface from the \textsf{R-INLA} \citep{RINLA_r} software. We compare the predictive performance of our stacking algorithm with MCMC by computing the mean log point-wise predictive density (MLPD) for the locations held-out, given by $\mathrm{MLPD}_{\text{stacking}} = n_h^{-1} \sum_{\ell \in \calL_h} \log \sum_{g = 1}^G \hat{w}_g p(y(\ell) \given y, M_g)$, and $\mathrm{MLPD}_{\text{MCMC}} = n_h^{-1} \sum_{\ell \in \calL_h} \log p(y(\ell) \given y)$. For INLA, we obtain an approximate MLPD $\hat{p}(y(\ell) \given y)$ using samples from the approximated posterior distribution.

\subsection{Results}
\subsubsection{Posterior learning and predictive performance}
\begin{figure}[t]
\centering
\includegraphics[width=0.7\linewidth]{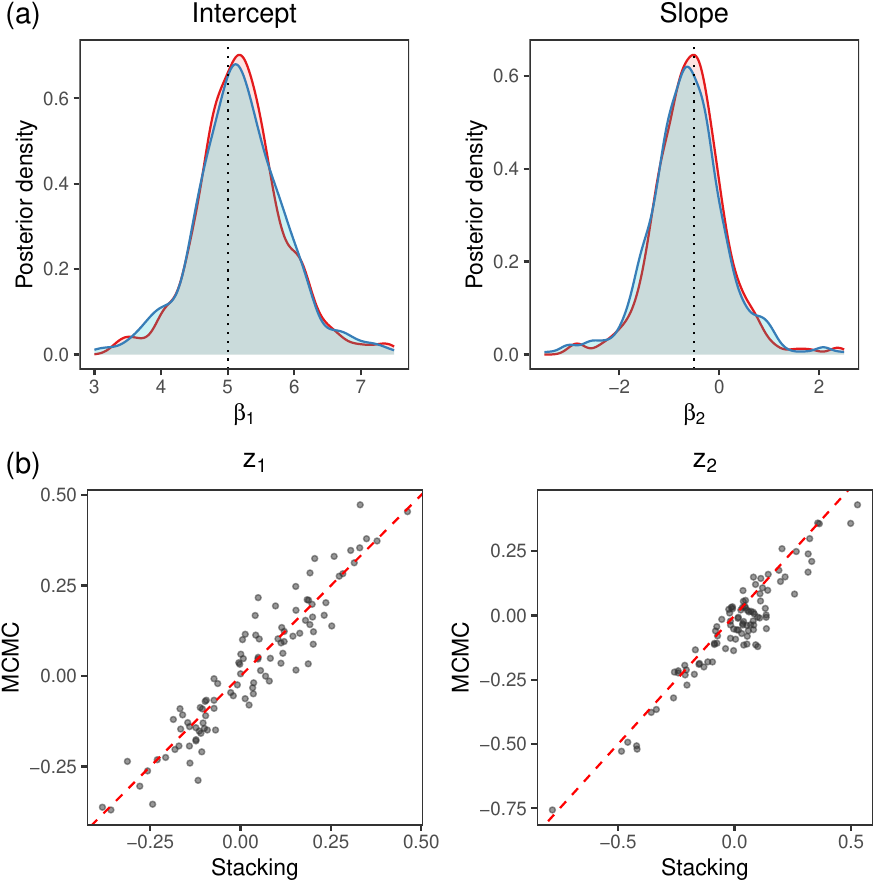}
\caption{The multivariate spatial-temporal process model fitted on simulated Poisson count data: (a) posterior distributions of the intercept as well as the slope obtained from stacking (blue) and MCMC (red) overlaid dotted vertical line showing their true values; (b) posterior medians of spatial-temporal random effects obtained by stacking and MCMC with the $y=x$ reference as a red dashed line.}
\label{fig:sptv_stack_vs_mcmc}
\end{figure}
Figures~\ref{fig:sptv_stack_vs_mcmc}(a)~and~(b) show results for the multivariate process model fitted to the simulated Poisson count data. Figure~\ref{fig:sptv_stack_vs_mcmc}(a) presents the overlaid posterior densities of the fixed effects obtained from the stacked posterior (blue) and MCMC (red), revealing practically indistinguishable posterior distributions. Figure~\ref{fig:sptv_stack_vs_mcmc}(b) also displays a high agreement between predictive stacking and MCMC with respect to the posterior medians of the spatial-temporal random effects associated with the intercept ($z_1$) and slope $z_2$. We note similar observations for the independent process model. We also notice that the posterior samples of the spatial-temporal process parameters do not necessarily concentrate around their true values, hence, demonstrating their weak identifiability. We observe similar patterns in the posterior distributions of the temporal ($\phi_{11}$ and $\phi_{12}$) and spatial decay parameters ($\phi_{21}$ and $\phi_{22}$) for the independent process model (Figure~\ref{fig:plot_phi} in the Appendix).

\begin{table}[t]
\centering
\resizebox{0.95\textwidth}{!}{%
{\begin{tabular}{@{}llcccccc@{}}
\toprule
\multirow{2}{*}{\begin{tabular}[c]{@{}c@{}}Data generative\\ process model \end{tabular}} & \multirow{2}{*}{Model} & \multirow{2}{*}{Method} & \multicolumn{5}{c}{Sample size ($n$)} \\ \cmidrule{4-8} 
&  & & $n = 100$   & $n = 200$  & $n = 300$  & $n = 400$ & $n = 500$ \\ \cmidrule{1-8}
\multirow{4}{*}{Multivariate} & \multirow{2}{*}{Multivariate}  & Stacking & -3.67  & -3.62   & -3.55   & -3.45  & -3.39  \\
& & MCMC &   -3.46  &  -3.44    &  -3.37    &  -3.36   &  -3.34 \\ 
& \multirow{2}{*}{Independent} & Stacking & -3.81  &  -3.75  &  -3.68  & -3.57  &  -3.49 \\
& & MCMC &  -3.58   &  -3.49  &  -3.41  &  -3.32  &  -3.27 \\ 
\cmidrule{1-8}
\multirow{4}{*}{Independent} & \multirow{2}{*}{Multivariate}  & Stacking & -3.65  & -3.59   & -3.49   & -3.41  & -3.29  \\
& & MCMC &   -3.58  &  -3.51   &  -3.42    &  -3.30   &  -3.20 \\ 
& \multirow{2}{*}{Independent} & Stacking & -3.79  &  -3.76  &  -3.72  & -3.52  &  -3.38\\
& & MCMC & -3.53   &  -3.46  &  -3.34  &  -3.22  &  -3.14 \\ 
\bottomrule
\end{tabular}%
}
}
\caption{Predictive performance of stacking and MCMC under correct and misspecified spatial-temporal models on simulated Poisson counts. All values correspond to mean log-pointwise predictive density (MLPD) at 100 held-out locations, based on 5 replications.}
\label{tab:mlpd}
\end{table}
Treating the fully Bayesian model with priors on $\thetasp$, which is fitted using MCMC, we find that the predictive performance of our proposed stacking algorithm becomes closer to MCMC as the sample size increases (Table~\ref{tab:mlpd}). For example, in the case of the multivariate model fitted on simulated data with multivariate process, we see that the difference in mean log-pointwise predictive density of 100 held-out samples between MCMC and our proposed stacking algorithm drops from 6.1\% at sample size 100 to 1.5\% at sample size 500. Table~\ref{tab:mlpd} reveals that the predictive precision of the multivariate model is better than the independent process model even when the data are simulated from the latter. This is possibly due to -- (i) the weak identifiability of the process parameters in the independent model, which has $2r$ process parameters compared to just $2$ in the multivariate model, and (ii) the random selection of candidate values for the $2r$ process parameters in the independent model against Cartesian product over a grid of candidate values of 2 process parameters in the multivariate model. In summary, the independent process model does not necessarily deliver superior predictive inference compared to the multivariate model in the presence of weakly identified parameters. We observe similar trends in the simulated binomial count and binary data (see Tables~\ref{tab:mlpd2_supp}~and~\ref{tab:mlpd3_supp} in the Appendix).

Moreover, we compare predictive stacking and INLA for a spatially-temporally varying coefficient Poisson regression model by evaluating predictive performance using MLPD and examining computational runtimes, highlighting the trade-off between accuracy and efficiency. We summarize our findings in Table~\ref{tab:inlavstack}.
\begin{table}[t]
\centering
\begin{tabular}{@{}rcccccc@{}}
\toprule
\multirow{2}{*}{} & \multirow{2}{*}{\begin{tabular}[c]{@{}c@{}}Sample\\ size \end{tabular}} & \multirow{2}{*}{\begin{tabular}[c]{@{}c@{}}Predictive\\ stacking \end{tabular}} &  & \multicolumn{3}{c}{INLA ($s_{\text{grid}}$, $t_{\text{grid}}$)} \\ \cmidrule(l){5-7} 
 &  &  & & (0.5, 0.2) & (0.25, 0.1) & (0.15, 0.1) \\ \midrule
\multirow{4}{*}{MLPD} 
 & 100 & -3.61 & & -3.87 & -3.94 & -3.96  \\
 & 200 & -3.53 & & -3.82 & -3.89 & -3.85  \\
 & 300 & -3.51 & & -3.78 & -3.79 & -3.64  \\
 & 400 & -3.46 & & -3.75 & -3.65 & -3.45  \\ \midrule
\multirow{4}{*}{Runtime} 
 & 100 & 2.9 & & 8.7 & 35.5 & 138.2  \\
 & 200 & 9.9 & & 8.8 & 35.5 & 139.1  \\
 & 300 & 31.8 & & 8.8 & 36.6 & 149.6  \\
 & 400 & 61.5 & & 8.9 & 36.9 & 153.2  \\ \bottomrule
\end{tabular}
\caption{Comparison of predictive performance and execution times between predictive stacking and INLA under different mesh resolutions on spatial-temporal point-referenced Poisson count data with varying sample sizes.}
\label{tab:inlavstack}
\end{table}
In our INLA experiments, $s_{\text{grid}}$ denotes the maximum edge length in the two-dimensional spatial mesh (see Figure~\ref{fig:inla-mesh} in the Appendix), and $t_{\text{grid}}$ denotes the length between equispaced temporal grid points. We observe that predictive stacking consistently offers superior predictive performance compared to INLA, even when the latter is fitted with finer mesh resolutions. Although finer meshes provide a more detailed discretization of the spatial-temporal domain, predictive performance does not necessarily improve with increased resolution. This behaviour is largely an artefact of analysing irregularly spaced data within the INLA framework, as finer meshes introduce additional nodes in regions without observations. In these areas, the latent field is poorly constrained, and INLA relies more heavily on the the smoothness assumption rather than the data, which can inflate uncertainty and may degrade predictive performance. Moreover, INLA’s reliance on Laplace approximation to handle high-dimensional latent processes can limit the flexibility of the inference, especially in models with multiple spatial-temporal processes, such as our spatially-temporally varying coefficient model. Here, it is also important to recognize that our predictive stacking approach and INLA differ fundamentally, since the former aggregates predictive distributions across candidate models, whereas the latter delivers inference by approximating the posterior distribution.

\subsubsection{Runtime comparison}
\begin{figure}[t]
    \centering
    \includegraphics[width=0.7\linewidth]{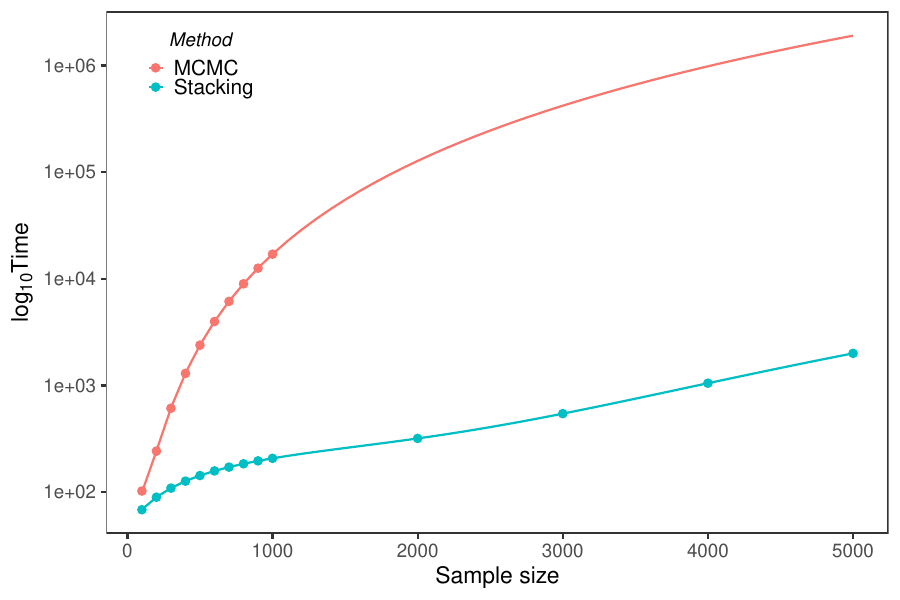}
    \caption{Comparison of runtimes of our proposed stacking algorithm and MCMC under the multivariate process assumption. Stacking demonstrates significantly faster execution times across various sample sizes compared to the MCMC algorithms.}
    \label{fig:runtime}
\end{figure}

Figure~\ref{fig:runtime} reveals that predictive stacking is, on average, about 500 times faster than MCMC further corroborating the efficiency of predictive stacking as an alternative to MCMC. We have implemented our predictive stacking algorithm within the \textsf{R} statistical computing environment (\url{https://www.r-project.org/}), where the code was written in \textsf{C++} and \textsf{Fortran} leveraging \textsf{BLAS} and \textsf{LAPACK} libraries for efficient matrix operations. We execute the programs for runtime comparisons on hardware equipped with an 1-Intel(R) Xeon(R) Gold 6140 CPU @ 2.30GHz processor with 36 cores and 1 thread per core, totalling 36 possible threads for parallel computing. The runtime for predictive stacking reported here is based on execution using 6 cores only, in order to reflect realistic computational resources in contemporary personal computers. We compare the runtime of our proposed algorithm with the adaptive Metropolis-within-Gibbs algorithm that implements full Bayesian inference.
Earlier benchmarks for MCMC-based inference for spatial GLMs, such as those reported by \textsf{spBayes}, indicate that computational demands become impractical when the sample size $n \sim 10^3$, or higher. In comparison, our proposed method achieves reasonable execution times for datasets with $n \sim 10^4$.

\section{North American Breeding Bird Survey}\label{sec:data_analysis}
\noindent Terrestrial birds were sampled annually on routes (approximately 40.23~km with point counts every 0.8~km) throughout the United States and Canada as part of the North American Breeding Bird Survey \citep{bbs}. We analyse the number of migrant birds observed at different locations across the United States with the goal of predicting their numbers at an arbitrary location. 

Our key inferential objective is to evaluate how the impact of nearby vehicles and external noise varies over space and time. We used 2,396 spatial coordinates between 2010 and 2019 at which the number of migrant birds were recorded. We aggregated the count data of different species within each spatial unit, since very few locations recorded counts for more than one species. We consider two explanatory variables, ``car'' and ``noise''. The former represents tallies of vehicles passing survey points during each 3-minute count, and the latter reports unrelated excessive noise at each point from sources other than passing vehicles (for example, from construction work). The presence of excessive noise is defined as noise lasting more than 45 seconds that significantly interferes with the observer's ability to hear birds at the location during the sampling period. These are mapped to the same GPS coordinates in the form of latitude and longitude that reference the avian counts.

The existing analysis has focused on modelling species-level variability and does not account for spatial-temporal correlation. For example, \cite{sauerlinkbbs1997} and \cite{sauerlinkbbs2011} study the route-level population trajectories of a species and observer effects using quasi-likelihood approaches and a Bayesian hierarchical log-linear model, respectively. \cite{sauerlinkbbs2011} elaborates on the practical difficulties arising from convergence issues that require a limited number of MCMC iterations for moderately large datasets. Furthermore, their model featured random effects of a much lower dimension than we have in our proposed model \eqref{eq:model_final}. In addition, none of the previous analyses uses spatial random effects in order to model spatial variability in their log-linear model, whereas we utilise available geographic coordinates of the routes to build spatial-temporal processes to account for  dependencies in the avian point count data within a rich modelling framework. This framework enables us to account for large scale variation in the mean using explanatory variables (for example the presence of excessive noise from vehicles) so the latent process evinces the residual spatial-temporal association in bird counts that can indicate lurking factors affecting the avian population.

We apply a spatial-temporal Poisson regression model in \eqref{eq:model_final} with the multivariate spatial-temporal process to analyse the counts of migrant birds. The predictors `car' and `noise' have spatially-temporally varying regression coefficients with the correlation function in \eqref{eq:corr_fn}. We assumed $\Psi = 1.5 I_3 + 0.51_31_3^\T$. We implement the proposed stacking with $G_{\phi_2} = \{40, 800, 1000 \}$ (the effective spatial range is approximately 20\%, 50\% and 70\% of the maximum distance between sites), $G_{\phi_1} = \{0.5, 1, 2\}$ and $G_{\alpha_\epsilon} = \{0.5, 0.75\}$. We also fit a non-spatial-temporal Poisson regression model and a spatial-temporal model with the intercept as the only predictor with varying coefficients, with the latter using our predictive stacking framework.
\begin{table}[t]
\centering
{\begin{tabular}{@{}lccc@{}}
\toprule
\multicolumn{1}{c}{\multirow{3}{*}{\begin{tabular}[c]{@{}c@{}}Model\\ parameters\end{tabular}}} & \multicolumn{3}{c}{Model} \\ \cmidrule(l){2-4} 
\multicolumn{1}{c}{} & \begin{tabular}[c]{@{}c@{}}Non-spatial-\\ temporal\end{tabular} & \begin{tabular}[c]{@{}c@{}}Intercept-only\\ spatial-temporal\end{tabular} & \begin{tabular}[c]{@{}c@{}}Spatially-temporally\\ varying coefficients\end{tabular} \\ \midrule
Intercept & \begin{tabular}[c]{@{}c@{}}2.182\\ (2.165, 2.195)\end{tabular} & \begin{tabular}[c]{@{}c@{}}0.946\\ (0.832, 1.063)\end{tabular} & \begin{tabular}[c]{@{}c@{}}0.874\\ (0.729, 1.109)\end{tabular} \\ \cmidrule{1-4}
Car & \begin{tabular}[c]{@{}c@{}}0.0003\\ (0.0002, 0.0004)\end{tabular} & \begin{tabular}[c]{@{}c@{}}0.0002\\ (-0.0004, 0.0008)\end{tabular} & \begin{tabular}[c]{@{}c@{}}0.003\\ (-0.021, 0.029)\end{tabular} \\ \cmidrule{1-4}
Noise & \begin{tabular}[c]{@{}c@{}}0.026\\ (0.023, 0.028)\end{tabular} & \begin{tabular}[c]{@{}c@{}}0.0121\\ (-0.007, 0.032)\end{tabular} & \begin{tabular}[c]{@{}c@{}}0.007\\ (-0.104,  0.118)\end{tabular} \\ \cmidrule{1-4}
Average bird count & \begin{tabular}[c]{@{}c@{}}9.749\\ (9.572, 9.819)\end{tabular} & \begin{tabular}[c]{@{}c@{}}5.136\\ (4.099, 10.984)\end{tabular} & \begin{tabular}[c]{@{}c@{}}9.472\\ (9.455, 10.218)\end{tabular} \\ \bottomrule
\end{tabular}%
}
\caption{North American Breeding Bird Survey (2010-19): Posterior summary of global regression coefficients and average bird count over all spatial-temporal coordinates. Values in parenthesis represent 2.5\% and 97.5\% quantiles.}
\label{tab:bbs_beta}
\end{table}
Table~\ref{tab:bbs_beta} presents posterior summaries of $\beta$ (the global regression coefficients that does not vary over space and time) for the intercept, `car' and `noise'. The global intercept is significantly larger than zero, which contributes approximately 2.4 units to the count with 95\% credible interval (2.1, 2.8) in the presence of no passing cars or excessive noise. Neither the number of cars nor levels of excessive noise seem to significantly impact the count in terms of global effects. The last row of Table~\ref{tab:bbs_beta} presents the estimated bird count averaged over all observed spatial-temporal coordinates in the data set. This is obtained from the posterior predictive distribution of the average bird count and provides us with an estimate of the relative influence of the spatially-temporally varying component over the global effects. Here, we see that the spatially-temporally varying coefficients increase the estimate of counts ($\approx$ 9.7) by a factor of 4 over the estimated bird count of $\approx 2.4$ estimated from the global effects. This estimate of the average bird count is consistent with what one would obtain from customary GLMs with fixed effects only, but simple GLMs will not offer insights into the spatially and temporally varying impact of predictors. Also, the intercept-only spatial-temporal model is underestimating the response, which may be due to excessive smoothing, where the random effects are not capturing local variations properly. This suggests the importance of accounting for spatial-temporal variations in regression slopes to accurately model the response.

\begin{figure}[t]
    \centering
    \includegraphics[width=0.9\linewidth]{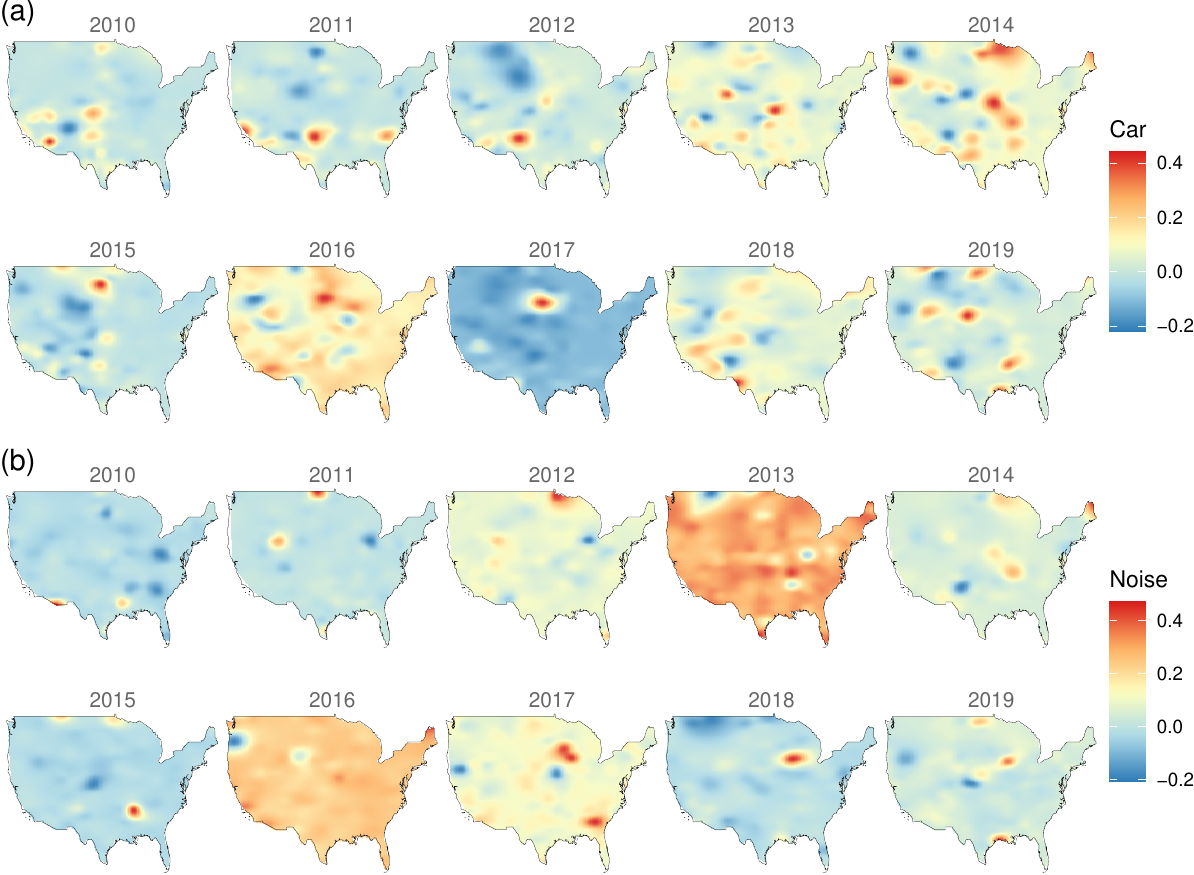}
    \caption{North American Breeding Bird Survey (2010-19): Interpolated surfaces of posterior median of spatial-temporal random effects in the slope of the variables (a) `car' and, (b) `noise' obtained by our proposed stacking algorithm.}
    \label{fig:bbs}
\end{figure}

Figures~\ref{fig:bbs}(a)~and~(b) reveal the space-time varying impact of the predictors `car' and `noise'. In Fig.~\ref{fig:bbs}(a) we see significant positive impact (red) of `car' in the Niland and Palo Verde regions of Imperial county, California, (in the south-west corner) rather consistently between 2010 and 2012. These elevated spatial-temporal coefficients and higher numbers of cars in these regions (not shown) produce higher than average estimates of bird counts there. We observe a similar pattern in parts of North Dakota and South Dakota between 2015 and 2019. Figure~\ref{fig:bbs}(b) reveals a positive impact (red) of `noise' in the spatial-temporal random effects in northern Minnesota from 2011 to 2015. While this area experiences persistently low levels of `noise' (not shown), the high values of the coefficients produce higher estimates of bird counts. These spatially-temporally varying coefficients capture the local impact of predictors to adjust the global effects from Table~\ref{tab:bbs_beta}. Appendix~\ref{sec:data_analysis_supp} presents additional data analysis.
Lastly, while predictive stacking delivered inference in 20 minutes using 6 cores, an adaptive Metropolis-within-Gibbs algorithm anticipates at least 30,000 iterations for convergence with each iteration taking 20 seconds on average. Hence, stacking offers speed-ups of over 500 times over MCMC.

\section{Discussion}\label{sec:discussion}
We developed Bayesian predictive stacking as an effective tool to estimate spatially-temporally varying regression coefficients and carry out predictive inference on non-Gaussian spatial-temporal data. Our method effectively combines inference across different closed-form posterior distributions by circumventing inference on weakly identified parameters. Rather than a replacement or competitor to MCMC, which, in principle, offers full uncertainty quantification of weakly identified parameters, we see stacking as an extremely viable tool in the spatial analysts' armour when primary inference concerns the process itself, regression coefficients, variance components and predictions. Our proposed stacking framework can be explored further in at least two directions of active research in scalable spatial analysis. First, we note that the current article has dealt with full Gaussian process models, which involves the Cholesky factorization of a dense covariance matrix incurring a computational cost of $\sim O(n^3)$ flops. This can be significantly mitigated by replacing the full Gaussian process with processes that can scale inference to massive amounts of data \citep[see. e.g.,][and references therein for a burgeoning literature on scalable Gaussian processes]{datta_nngp_2016, sbBA2017, heaton2019_casestudy_jabes, peruzziEtAl2022jasa}. Here, we do not anticipate substantial modifications since scalable processes retain similar parametric forms and conjugacy of conditional posteriors \citep[see, e.g.,][]{datta_nngp_2016, zhangEtAl2019sam, banerjee_massivespatial} and stacking can be executed over a grid of their covariance kernel parameters as described here. A different avenue for scalable stacking is federated Bayesian learning and transfer learning, where inference is carried out on distributed data subsets \citep[see, e.g.,][]{guhaniyogiBanerjee2018technometrics, vehtariEtAl2020jmlr, guhaniyogiEtAl2023StatSci} and the resulting subset posteriors are combined via stacking \citep{presicce2025bayesiantransferlearningartificially} providing a complementary route to computational scalability. 

A second direction of research under exploration attempts to address what we acknowledge is a limitation of the current methodology. Note that predictive stacking relies on a discrete grid over a subset of model parameters. If the number of such parameters (typically hyperparameters in covariance kernels) in the model is very large, then computation becomes impractical because the number of candidate models grows exponentially with the dimension of the grid. For most spatial and spatial-temporal models, this issue seldom arises in practice, as the number of weakly identified parameters requiring conditioning is typically small. However in other settings \citep[e.g., highly multivariate responses][]{zhangbanerjee2021, zhangBanerjee2022biocs, ggp_dey2022, peruzziEtAl2025ba}, where we seek to jointly model multiple spatial processes, hyperparameters in the resulting spatial cross-covariance functions can become high-dimensional and our current method becomes computationally infeasible. While some mitigation of this problem is achieved by using a random sample of values of these parameters (rather than every lattice point on the high-dimensional grid), there is ample scope for research in this domain. One promising direction is the development of methodologies that reduce the effective number of candidate models through principled compression of the parameter space, while preserving the inferential objectives of stacking. Such approaches may offer a pathway to both computational scalability and improved flexibility in high-dimensional settings. Finally, it remains an open problem to investigate the effectiveness of predictive stacking for non-stationary processes with spatially-varying covariance kernels \citep{paciorek06}. In these models covariance kernel parameters are themselves modelled as spatial processes, so conditioning on their realisations is impractical. However, recent developments in \cite{coubeEtAl2025jcgs}, which use scalable processes to model covariance kernel parameters to achieve dimension reduction, can offer promising avenues of research to adapt such models to be analysed using predictive stacking.

\section*{Acknowledgement}
\noindent This work used computational and storage services associated with the Hoffman2 Shared Cluster provided by UCLA Office of Advanced Research Computing's Research Technology Group. Sudipto Banerjee and Soumyakanti Pan were supported by two research grants from the National Institute of Environmental Health Sciences (NIEHS), one grant from the National Institute of General Medical Science (NIGMS) and another from the Division of Mathematical Sciences of the National Science Foundation (NSF-DMS). Lu Zhang was supported by NIEHS (P30ES007048, P20HL176204, R01ES031590). J. R. Bradley was supported by NSF-DMS.

\section*{Data and code availability}
\noindent All code and data required to reproduce the results and findings in this article are openly available at \url{https://github.com/SPan-18/stvcGLMstack}. The code is disseminated via the development version of the R package \textsf{spStack} \citep{spStack_r}, available for download from the GitHub repository \url{https://github.com/SPan-18/spStack-dev}. The North American Breeding Bird survey database is openly available at \url{https://doi.org/10.5066/P97WAZE5}.


\bibliographystyle{plainnat}
\bibliography{refs}

\begin{thebibliography}{61}
\providecommand{\natexlab}[1]{#1}
\providecommand{\url}[1]{\texttt{#1}}
\expandafter\ifx\csname urlstyle\endcsname\relax
  \providecommand{\doi}[1]{doi: #1}\else
  \providecommand{\doi}{doi: \begingroup \urlstyle{rm}\Url}\fi

\bibitem[Abramowitz and Stegun(1965)]{abramowitzstegun}
Milton Abramowitz and Irene~A. Stegun, editors.
\newblock \emph{Handbook of Mathematical Functions with Formulas, Graphs and
  Mathematical Tables}.
\newblock Dover Publications, Inc., New York, 1965.

\bibitem[ApS(2023)]{mosek}
MOSEK ApS.
\newblock \emph{MOSEK Rmosek package 9.3.22}, 2023.
\newblock URL \url{https://docs.mosek.com/9.3/rmosek/index.html}.

\bibitem[Banerjee(2017)]{sbBA2017}
Sudipto Banerjee.
\newblock {High-Dimensional {B}ayesian Geostatistics}.
\newblock \emph{{B}ayesian Analysis}, 12\penalty0 (2):\penalty0 583 -- 614,
  2017.
\newblock \doi{10.1214/17-BA1056R}.
\newblock URL \url{https://doi.org/10.1214/17-BA1056R}.

\bibitem[Banerjee(2020)]{banerjee_massivespatial}
Sudipto Banerjee.
\newblock Modeling massive spatial datasets using a conjugate {B}ayesian linear
  modeling framework.
\newblock \emph{Spatial Statistics}, 37:\penalty0 100417, 2020.
\newblock ISSN 2211-6753.
\newblock \doi{10.1016/j.spasta.2020.100417}.
\newblock URL \url{https://doi.org/10.1016/j.spasta.2020.100417}.
\newblock Frontiers in Spatial and Spatio-temporal Research.

\bibitem[Banerjee et~al.(2025)Banerjee, Gelfand, and Carlin]{banerjee_spatial}
Sudipto Banerjee, Alan~E. Gelfand, and Brad~P. Carlin.
\newblock \emph{Hierarchical Modeling and Analysis for Spatial Data}.
\newblock Chapman and Hall/CRC, 3rd edition, 2025.
\newblock \doi{10.1201/9781003401728}.
\newblock URL \url{https://doi.org/10.1201/9781003401728}.

\bibitem[Bradley and Clinch(2024)]{bradley2023lgp}
Jonathan~R. Bradley and Madelyn Clinch.
\newblock Generating independent replicates directly from the posterior
  distribution for a class of spatial hierarchical models.
\newblock \emph{Journal of Computational and Graphical Statistics}, 0\penalty0
  (ja):\penalty0 1--32, 2024.
\newblock \doi{10.1080/10618600.2024.2365728}.
\newblock URL \url{https://doi.org/10.1080/10618600.2024.2365728}.

\bibitem[Bradley et~al.(2020)Bradley, Holan, and Wikle]{bradleyholanwikle}
Jonathan~R. Bradley, Scott~H. Holan, and Christopher~K. Wikle.
\newblock {B}ayesian hierarchical models with conjugate full-conditional
  distributions for dependent data from the natural exponential family.
\newblock \emph{Journal of the American Statistical Association}, 115\penalty0
  (532):\penalty0 2037--2052, 2020.
\newblock \doi{10.1080/01621459.2019.1677471}.
\newblock URL \url{https://doi.org/10.1080/01621459.2019.1677471}.

\bibitem[Breiman(1996)]{breiman1996stacked}
Leo Breiman.
\newblock Stacked regressions.
\newblock \emph{Machine learning}, 24\penalty0 (1):\penalty0 49--64, 1996.

\bibitem[Clyde and Iversen(2013)]{clyde2013bayesian}
Merlise Clyde and Edwin~S Iversen.
\newblock {B}ayesian model averaging in the m-open framework.
\newblock \emph{{B}ayesian theory and applications}, 14\penalty0 (4):\penalty0
  483--498, 2013.
\newblock \doi{10.1093/acprof:oso/9780199695607.003.0024}.
\newblock URL \url{https://doi.org/10.1093/acprof:oso/9780199695607.003.0024}.

\bibitem[Coube-Sisqueille et~al.(2025)Coube-Sisqueille, Banerjee, and
  Liquet]{coubeEtAl2025jcgs}
Sébastien Coube-Sisqueille, Sudipto Banerjee, and Benoît Liquet.
\newblock Nonstationary spatial process models with spatially varying
  covariance kernels.
\newblock \emph{Journal of Computational and Graphical Statistics}, (in press),
  2025.
\newblock \doi{10.1080/10618600.2025.2516020}.
\newblock URL \url{https://doi.org/10.1080/10618600.2025.2516020}.

\bibitem[Datta et~al.(2016)Datta, Banerjee, Finley, and
  Gelfand]{datta_nngp_2016}
Abhirup Datta, Sudipto Banerjee, Andrew~O. Finley, and Alan~E. Gelfand.
\newblock {Hierarchical Nearest-Neighbor Gaussian Process Models for Large
  Geostatistical Datasets}.
\newblock \emph{Journal of the American Statistical Association}, 111\penalty0
  (514):\penalty0 800--812, 2016.
\newblock \doi{10.1080/01621459.2015.1044091}.
\newblock URL \url{https://doi.org/10.1080/01621459.2015.1044091}.

\bibitem[{De Oliveira}(2000)]{de2000bayesian}
Victor {De Oliveira}.
\newblock {B}ayesian prediction of clipped gaussian random fields.
\newblock \emph{Computational Statistics \& Data Analysis}, 34\penalty0
  (3):\penalty0 299--314, 2000.
\newblock \doi{10.1016/S0167-9473(99)00103-6}.
\newblock URL \url{https://doi.org/10.1016/S0167-9473(99)00103-6}.

\bibitem[{De Oliveira} et~al.(1997){De Oliveira}, Kedem, and
  Short]{deoliveira1997tgrf}
Victor {De Oliveira}, Benjamin Kedem, and David~A. Short.
\newblock {B}ayesian prediction of transformed gaussian random fields.
\newblock \emph{Journal of the American Statistical Association}, 92\penalty0
  (440):\penalty0 1422--1433, 1997.
\newblock \doi{10.1080/01621459.1997.10473663}.
\newblock URL \url{https://doi.org/10.1080/01621459.1997.10473663}.

\bibitem[Dey et~al.(2022)Dey, Datta, and Banerjee]{ggp_dey2022}
Debangan Dey, Abhirup Datta, and Sudipto Banerjee.
\newblock Graphical gaussian process models for highly multivariate spatial
  data.
\newblock \emph{Biometrika}, 109\penalty0 (4):\penalty0 993--1014, 12 2022.
\newblock ISSN 1464-3510.
\newblock \doi{10.1093/biomet/asab061}.

\bibitem[Diaconis and Ylvisaker(1979)]{DY79}
Persi Diaconis and Donald Ylvisaker.
\newblock {Conjugate Priors for Exponential Families}.
\newblock \emph{The Annals of Statistics}, 7\penalty0 (2):\penalty0 269 -- 281,
  1979.
\newblock \doi{10.1214/aos/1176344611}.
\newblock URL \url{https://doi.org/10.1214/aos/1176344611}.

\bibitem[Diggle et~al.(1998)Diggle, Tawn, and Moyeed]{diggle_geostat_1998}
P.~J. Diggle, J.~A. Tawn, and R.~A. Moyeed.
\newblock Model-based geostatistics.
\newblock \emph{Journal of the Royal Statistical Society Series C: Applied
  Statistics}, 47\penalty0 (3):\penalty0 299--350, 01 1998.
\newblock ISSN 0035-9254.
\newblock \doi{10.1111/1467-9876.00113}.
\newblock URL \url{https://doi.org/10.1111/1467-9876.00113}.

\bibitem[Ding(2016)]{ding2016conditional}
Peng Ding.
\newblock On the conditional distribution of the multivariate t distribution.
\newblock \emph{The American Statistician}, 70\penalty0 (3):\penalty0 293--295,
  2016.
\newblock \doi{10.1080/00031305.2016.1164756}.
\newblock URL \url{https://doi.org/10.1080/00031305.2016.1164756}.

\bibitem[Finley et~al.(2015)Finley, Banerjee, and Gelfand]{spBayes_r}
Andrew~O. Finley, Sudipto Banerjee, and Alan~E. Gelfand.
\newblock spbayes for large univariate and multivariate point-referenced
  spatio-temporal data models.
\newblock \emph{Journal of Statistical Software}, 63\penalty0 (13):\penalty0
  1–28, 2015.
\newblock \doi{10.18637/jss.v063.i13}.
\newblock URL \url{https://doi.org/10.18637/jss.v063.i13}.

\bibitem[Fu and Narasimhan(2023)]{ecos}
Anqi Fu and Balasubramanian Narasimhan.
\newblock \emph{ECOSolveR: Embedded Conic Solver in R}, 2023.
\newblock URL \url{https://bnaras.github.io/ECOSolveR/}.
\newblock R package version 0.5.5.

\bibitem[Fu et~al.(2020)Fu, Narasimhan, and Boyd]{cvxr2020}
Anqi Fu, Balasubramanian Narasimhan, and Stephen Boyd.
\newblock {CVXR}: An {R} package for disciplined convex optimization.
\newblock \emph{Journal of Statistical Software}, 94\penalty0 (14):\penalty0
  1--34, 2020.
\newblock \doi{10.18637/jss.v094.i14}.
\newblock URL \url{https://doi.org/10.18637/jss.v094.i14}.

\bibitem[Gelfand et~al.(2003)Gelfand, Kim, Sirmans, and
  Banerjee]{gelfand2003svc}
Alan~E. Gelfand, Hyon-Jung Kim, C.~F. Sirmans, and Sudipto Banerjee.
\newblock Spatial modeling with spatially varying coefficient processes.
\newblock \emph{Journal of the American Statistical Association}, 98\penalty0
  (462):\penalty0 387--396, 2003.
\newblock \doi{10.1198/016214503000170}.
\newblock URL \url{https://doi.org/10.1198/016214503000170}.

\bibitem[Gneiting and Guttorp(2010)]{gnei10}
T.~Gneiting and P.~Guttorp.
\newblock Continuous-parameter spatio-temporal processes.
\newblock In A.E. Gelfand, P.J. Diggle, M.~Fuentes, and P~Guttorp, editors,
  \emph{Handbook of Spatial Statistics}, Chapman \& Hall CRC Handbooks of
  Modern Statistical Methods, pages 427--436. Taylor and Francis, 2010.
\newblock \doi{10.1201/9781420072884}.
\newblock URL \url{https://doi.org/10.1201/9781420072884}.

\bibitem[Golub and Van~Loan(2013)]{GolubLoanMatrix4}
Gene~H. Golub and Charles~F. Van~Loan.
\newblock \emph{Matrix Computations}.
\newblock Johns Hopkins University Press, Philadelphia, PA, 4th edition, 2013.
\newblock \doi{10.1137/1.9781421407944}.
\newblock URL \url{https://doi.org/10.1137/1.9781421407944}.

\bibitem[Guhaniyogi and Banerjee(2018)]{guhaniyogiBanerjee2018technometrics}
Rajarshi Guhaniyogi and Sudipto Banerjee.
\newblock Meta-kriging: Scalable bayesian modeling and inference for massive
  spatial datasets.
\newblock \emph{Technometrics}, 60\penalty0 (4):\penalty0 430--444, 2018.
\newblock \doi{10.1080/00401706.2018.1437474}.
\newblock URL \url{https://doi.org/10.1080/00401706.2018.1437474}.

\bibitem[Guhaniyogi et~al.(2023)Guhaniyogi, Li, Savitsky, and
  Srivastava]{guhaniyogiEtAl2023StatSci}
Rajarshi Guhaniyogi, Cheng Li, Terrance Savitsky, and Sanvesh Srivastava.
\newblock {Distributed Bayesian Inference in Massive Spatial Data}.
\newblock \emph{Statistical Science}, 38\penalty0 (2):\penalty0 262 -- 284,
  2023.
\newblock \doi{10.1214/22-STS868}.
\newblock URL \url{https://doi.org/10.1214/22-STS868}.

\bibitem[Gupta and Nagar(1999)]{guptanagar_matrixvariate}
A.~K. Gupta and D.~K. Nagar.
\newblock \emph{Matrix Variate Distributions}.
\newblock Chapman and Hall/CRC, New York, 1999.
\newblock \doi{10.1201/9780203749289}.
\newblock URL \url{https://doi.org/10.1201/9780203749289}.

\bibitem[Haran(2011)]{haran2011}
M.~Haran.
\newblock Gaussian random field models for spatial data.
\newblock In \emph{Markov chain {M}onte {C}arlo Handbook Eds. Brooks, S.P.,
  Gelman, A.E. Jones, G.L. and Meng, X.L.}, pages 449--478. Chapman and
  {H}all/{CRC}, 2011.

\bibitem[Heagerty and Lele(1998)]{heagerty1998composite}
Patrick~J Heagerty and Subhash~R Lele.
\newblock A composite likelihood approach to binary spatial data.
\newblock \emph{Journal of the American Statistical Association}, 93\penalty0
  (443):\penalty0 1099--1111, 1998.
\newblock \doi{10.1080/01621459.1998.10473771}.
\newblock URL \url{https://doi.org/10.1080/01621459.1998.10473771}.

\bibitem[Heaton et~al.(2019)Heaton, Datta, Finley, Furrer, Guinness,
  Guhaniyogi, Gerber, Gramacy, Hammerling, Katzfuss, Lindgren, Nychka, Sun, and
  Zammit-Mangion]{heaton2019_casestudy_jabes}
Matthew~J. Heaton, Abhirup Datta, Andrew~O. Finley, Reinhard Furrer, Joseph
  Guinness, Rajarshi Guhaniyogi, Florian Gerber, Robert~B. Gramacy, Dorit
  Hammerling, Matthias Katzfuss, Finn Lindgren, Douglas~W. Nychka, Furong Sun,
  and Andrew Zammit-Mangion.
\newblock A case study competition among methods for analyzing large spatial
  data.
\newblock \emph{Journal of Agricultural, Biological, and Environmental
  Statistics}, 24\penalty0 (3):\penalty0 pp. 398--425, 2019.
\newblock \doi{10.1007/s13253-018-00348-w}.
\newblock URL \url{https://doi.org/10.1007/s13253-018-00348-w}.

\bibitem[Hoeting et~al.(1999)Hoeting, Madigan, Raftery, and
  Volinsky]{hoeting1999bma}
Jennifer~A. Hoeting, David Madigan, Adrian~E. Raftery, and Chris~T. Volinsky.
\newblock {{B}ayesian model averaging: a tutorial (with comments by M. Clyde,
  David Draper and E. I. George, and a rejoinder by the authors}.
\newblock \emph{Statistical Science}, 14\penalty0 (4):\penalty0 382 -- 417,
  1999.
\newblock \doi{10.1214/ss/1009212519}.
\newblock URL \url{https://doi.org/10.1214/ss/1009212519}.

\bibitem[Hughes and Haran(2013)]{hughes_haran_2013}
John Hughes and Murali Haran.
\newblock Dimension reduction and alleviation of confounding for spatial
  generalized linear mixed models.
\newblock \emph{Journal of the Royal Statistical Society. Series B: Statistical
  Methodology}, 75\penalty0 (1):\penalty0 139--159, January 2013.
\newblock \doi{10.1111/j.1467-9868.2012.01041.x}.
\newblock URL \url{https://doi.org/10.1111/j.1467-9868.2012.01041.x}.

\bibitem[Kim et~al.(2002)Kim, Park, and and]{kimcox2002_CV}
Tae~Yoon Kim, Jeong~Soo Park, and Dennis D~Cox and.
\newblock Fast algorithm for cross-validation of the best linear unbiased
  predictor.
\newblock \emph{Journal of Computational and Graphical Statistics}, 11\penalty0
  (4):\penalty0 823--835, 2002.
\newblock \doi{10.1198/106186002826}.
\newblock URL \url{https://doi.org/10.1198/106186002826}.

\bibitem[Krainski et~al.(2025)Krainski, Lindgren, and Rue]{INLAspacetime_r}
Elias~Teixeira Krainski, Finn Lindgren, and Haavard Rue.
\newblock \emph{INLAspacetime: Spatial and Spatio-Temporal Models using
  'INLA'}, 2025.
\newblock URL \url{https://CRAN.R-project.org/package=INLAspacetime}.
\newblock R package version 0.1.12.

\bibitem[Le and Clarke(2017)]{le2017bayes}
Tri Le and Bertrand Clarke.
\newblock A {B}ayes interpretation of stacking for m-complete and m-open
  settings.
\newblock \emph{{B}ayesian Analysis}, 12\penalty0 (3):\penalty0 807--829, 2017.
\newblock \doi{10.1214/16-BA1023}.
\newblock URL \url{https://doi.org/10.1214/16-BA1023}.

\bibitem[Lindgren and Rue(2015)]{RINLA_r}
Finn Lindgren and H{\aa}vard Rue.
\newblock Bayesian spatial modelling with {R}-{INLA}.
\newblock \emph{Journal of Statistical Software}, 63\penalty0 (19):\penalty0
  1--25, 2015.
\newblock \doi{10.18637/jss.v063.i19}.
\newblock URL \url{https://doi.org/10.18637/jss.v063.i19}.

\bibitem[Link and Sauer(1997)]{sauerlinkbbs1997}
William~A. Link and John~R. Sauer.
\newblock Estimation of population trajectories from count data.
\newblock \emph{Biometrics}, 53\penalty0 (2):\penalty0 488--497, 1997.
\newblock \doi{10.2307/2533952}.
\newblock URL \url{https://doi.org/10.2307/2533952}.

\bibitem[Madigan et~al.(1996)Madigan, Raftery, Volinsky, and
  Hoeting]{madigan1996bayesian}
David Madigan, Adrian~E Raftery, C~Volinsky, and Jennifer Hoeting.
\newblock {B}ayesian model averaging.
\newblock In \emph{Proceedings of the AAAI Workshop on Integrating Multiple
  Learned Models, Portland, OR}, pages 77--83, 1996.

\bibitem[Mardia and Goodall(1993)]{mardiagoodall1993}
K.~V. Mardia and C.~R. Goodall.
\newblock Spatial-temporal analysis of multivariate environmental monitoring
  data.
\newblock \emph{Multivariate Environmental Statistics}, 1993.

\bibitem[McCulloch and Searle(2001)]{glmm_mcculloch}
Charles~E. McCulloch and Shayle~R. Searle.
\newblock \emph{Generalized, linear, and mixed models}.
\newblock Wiley Series in Probability and Statistics. John Wiley \& Sons, 2001.
\newblock \doi{10.1002/0471722073}.
\newblock URL \url{https://doi.org/10.1002/0471722073}.

\bibitem[Paciorek and Schervish(2006)]{paciorek06}
Christopher~J. Paciorek and Mark~J. Schervish.
\newblock Spatial modelling using a new class of nonstationary covariance
  functions.
\newblock \emph{Environmetrics}, pages 483--506, 2006.
\newblock \doi{10.1002/env.785}.
\newblock URL \url{https://doi.org/10.1002/env.785}.

\bibitem[Pan and Banerjee(2024)]{spStack_r}
Soumyakanti Pan and Sudipto Banerjee.
\newblock \emph{{spStack}: {B}ayesian Geostatistics Using Predictive Stacking},
  2024.
\newblock URL \url{https://CRAN.R-project.org/package=spStack}.
\newblock R package version 1.1.2.

\bibitem[Peruzzi et~al.(2022)Peruzzi, Banerjee, and
  Finley]{peruzziEtAl2022jasa}
Michele Peruzzi, Sudipto Banerjee, and Andrew~O. Finley.
\newblock Highly scalable {B}ayesian geostatistical modeling via meshed
  {G}aussian processes on partitioned domains.
\newblock \emph{Journal of the American Statistical Association}, 117\penalty0
  (538):\penalty0 969--982, 2022.
\newblock \doi{10.1080/01621459.2020.1833889}.
\newblock URL \url{https://doi.org/10.1080/01621459.2020.1833889}.

\bibitem[Peruzzi et~al.(2025)Peruzzi, Banerjee, Dunson, and
  Finley]{peruzziEtAl2025ba}
Michele Peruzzi, Sudipto Banerjee, David~B. Dunson, and Andrew~O. Finley.
\newblock {Gridding and Parameter Expansion for Scalable Latent Gaussian Models
  of Spatial Multivariate Data}.
\newblock \emph{Bayesian Analysis}, pages 1 -- 27, 2025.
\newblock \doi{10.1214/25-BA1515}.
\newblock URL \url{https://doi.org/10.1214/25-BA1515}.

\bibitem[Presicce and
  Banerjee(2025)]{presicce2025bayesiantransferlearningartificially}
Luca Presicce and Sudipto Banerjee.
\newblock Bayesian transfer learning for artificially intelligent geospatial
  systems: A predictive stacking approach, 2025.
\newblock URL \url{https://arxiv.org/abs/2410.09504}.

\bibitem[Roberts and Rosenthal(2009)]{adaMetropGibbs_2009}
Gareth~O. Roberts and Jeffrey~S. Rosenthal.
\newblock Examples of adaptive {MCMC}.
\newblock \emph{Journal of Computational and Graphical Statistics}, 18\penalty0
  (2):\penalty0 349--367, 2009.
\newblock \doi{10.1198/jcgs.2009.06134}.
\newblock URL \url{https://doi.org/10.1198/jcgs.2009.06134}.

\bibitem[Rue et~al.(2009)Rue, Martino, and Chopin]{inla2009}
Håvard Rue, Sara Martino, and Nicolas Chopin.
\newblock {Approximate {B}ayesian Inference for Latent Gaussian models by using
  Integrated Nested Laplace Approximations}.
\newblock \emph{Journal of the Royal Statistical Society Series B: Statistical
  Methodology}, 71\penalty0 (2):\penalty0 319--392, 04 2009.
\newblock ISSN 1369-7412.
\newblock \doi{10.1111/j.1467-9868.2008.00700.x}.
\newblock URL \url{https://doi.org/10.1111/j.1467-9868.2008.00700.x}.

\bibitem[Saha et~al.(2022)Saha, Datta, and Banerjee]{saha_datta_banerjee_2022}
Arkajyoti Saha, Abhirup Datta, and Sudipto Banerjee.
\newblock Scalable predictions for spatial probit linear mixed models using
  nearest neighbor gaussian processes.
\newblock \emph{Journal of Data Science}, 20\penalty0 (4):\penalty0 533--544,
  2022.
\newblock ISSN 1680-743X.
\newblock \doi{10.6339/22-JDS1073}.
\newblock URL \url{https://doi.org/10.6339/22-JDS1073}.

\bibitem[Sauer and Link(2011)]{sauerlinkbbs2011}
John~R. Sauer and William~A. Link.
\newblock Analysis of the north american breeding bird survey using
  hierarchical models.
\newblock \emph{The Auk}, 128\penalty0 (1):\penalty0 87--98, 01 2011.
\newblock ISSN 1938-4254.
\newblock \doi{10.1525/auk.2010.09220}.
\newblock URL \url{https://doi.org/10.1525/auk.2010.09220}.

\bibitem[Vehtari and Lampinen(2002)]{vehtariCV2002}
Aki Vehtari and Jouko Lampinen.
\newblock {{B}ayesian Model Assessment and Comparison Using Cross-Validation
  Predictive Densities}.
\newblock \emph{Neural Computation}, 14\penalty0 (10):\penalty0 2439--2468, 10
  2002.
\newblock ISSN 0899-7667.
\newblock \doi{10.1162/08997660260293292}.
\newblock URL \url{https://doi.org/10.1162/08997660260293292}.

\bibitem[Vehtari et~al.(2017)Vehtari, Gelman, and Gabry]{vehtari_loo17}
Aki Vehtari, Andrew Gelman, and Jonah Gabry.
\newblock Practical {B}ayesian model evaluation using leave-one-out
  cross-validation and waic.
\newblock \emph{Statistics and Computing}, 27\penalty0 (5):\penalty0
  1413–1432, sep 2017.
\newblock ISSN 0960-3174.
\newblock \doi{10.1007/s11222-016-9696-4}.
\newblock URL \url{https://doi.org/10.1007/s11222-016-9696-4}.

\bibitem[Vehtari et~al.(2020)Vehtari, Gelman, Sivula, Jyl{\"a}nki, Tran, Sahai,
  Blomstedt, Cunningham, Schorfheide, and Yao]{vehtariEtAl2020jmlr}
Aki Vehtari, Andrew Gelman, Tuomas Sivula, Pasi Jyl{\"a}nki, Dustin Tran,
  Swupnil Sahai, Paul Blomstedt, John~P. Cunningham, Frank Schorfheide, and
  Yuling Yao.
\newblock A framework for bayesian inference on partitioned data.
\newblock \emph{Journal of Machine Learning Research}, 21\penalty0
  (87):\penalty0 1--49, 2020.
\newblock URL \url{http://jmlr.org/papers/v21/18-817.html}.

\bibitem[Wolpert(1992)]{wolpert1992stacked}
David~H Wolpert.
\newblock Stacked generalization.
\newblock \emph{Neural networks}, 5\penalty0 (2):\penalty0 241--259, 1992.

\bibitem[Yao et~al.(2018)Yao, Vehtari, Simpson, and Gelman]{yao2018using}
Yuling Yao, Aki Vehtari, Daniel Simpson, and Andrew Gelman.
\newblock Using stacking to average {B}ayesian predictive distributions (with
  discussion).
\newblock \emph{{B}ayesian Analysis}, 13\penalty0 (3):\penalty0 917--1007,
  2018.
\newblock \doi{10.1214/17-BA1091}.
\newblock URL \url{https://doi.org/10.1214/17-BA1091}.

\bibitem[Yao et~al.(2021)Yao, Pir\v{s}, Vehtari, and Gelman]{yao21_duplicate}
Yuling Yao, Gregor Pir\v{s}, Aki Vehtari, and Andrew Gelman.
\newblock {B}ayesian hierarchical stacking: {S}ome models are (somewhere)
  useful.
\newblock \emph{{B}ayesian Analysis}, 1\penalty0 (1):\penalty0 1--29, 2021.
\newblock \doi{10.1214/21-BA1287}.
\newblock URL \url{https://doi.org/10.1214/21-BA1287}.

\bibitem[Yao et~al.(2022)Yao, Vehtari, and Gelman]{yao2020stacking}
Yuling Yao, Aki Vehtari, and Andrew Gelman.
\newblock Stacking for {N}on-mixing {B}ayesian computations: {T}he curse and
  blessing of multimodal posteriors.
\newblock \emph{Journal of Machine Learning Research}, 23\penalty0
  (79):\penalty0 1--45, 2022.
\newblock URL \url{http://jmlr.org/papers/v23/20-1426.html}.

\bibitem[Zhang and Banerjee(2022)]{zhangBanerjee2022biocs}
Lu~Zhang and Sudipto Banerjee.
\newblock Spatial factor modeling: A bayesian matrix-normal approach for
  misaligned data.
\newblock \emph{Biometrics}, 78\penalty0 (2):\penalty0 560--573, 2022.
\newblock \doi{10.1111/biom.13452}.
\newblock URL \url{https://doi.org/10.1111/biom.13452}.

\bibitem[Zhang et~al.(2019)Zhang, Datta, and Banerjee]{zhangEtAl2019sam}
Lu~Zhang, Abhirup Datta, and Sudipto Banerjee.
\newblock Practical bayesian modeling and inference for massive spatial data
  sets on modest computing environments†.
\newblock \emph{Statistical Analysis and Data Mining: The ASA Data Science
  Journal}, 12\penalty0 (3):\penalty0 197--209, 2019.
\newblock \doi{10.1002/sam.11413}.
\newblock URL \url{https://doi.org/10.1002/sam.11413}.

\bibitem[Zhang et~al.(2021)Zhang, Banerjee, and Finley]{zhangbanerjee2021}
Lu~Zhang, Sudipto Banerjee, and Andrew~O. Finley.
\newblock High-dimensional multivariate geostatistics: A {B}ayesian
  matrix-normal approach.
\newblock \emph{Environmetrics}, 32\penalty0 (4):\penalty0 e2675, 2021.
\newblock \doi{10.1002/env.2675}.
\newblock URL \url{https://doi.org/10.1002/env.2675}.

\bibitem[Zhang et~al.(2025)Zhang, Tang, and Banerjee]{zhang2025jasa}
Lu~Zhang, Wenpin Tang, and Sudipto Banerjee.
\newblock Bayesian geostatistics using predictive stacking.
\newblock \emph{Journal of the American Statistical Association}, (In press),
  2025.
\newblock \doi{10.1080/01621459.2025.2566449}.
\newblock URL \url{https://doi.org/10.1080/01621459.2025.2566449}.

\bibitem[Zhang et~al.(2022)Zhang, Arellano-Valle, Genton, and
  Huser]{zhang2021tractable}
Zhongwei Zhang, Reinaldo~B. Arellano-Valle, Marc~G. Genton, and Raphaël Huser.
\newblock Tractable bayes of skew-elliptical link models for correlated binary
  data.
\newblock \emph{Biometrics}, 79\penalty0 (3):\penalty0 1788--1800, 08 2022.
\newblock ISSN 0006-341X.
\newblock \doi{10.1111/biom.13731}.
\newblock URL \url{https://doi.org/10.1111/biom.13731}.

\bibitem[Ziolkowski~Jr. et~al.(2022)Ziolkowski~Jr., Lutmerding, Aponte, and
  Hudson]{bbs}
David Ziolkowski~Jr., Michael Lutmerding, Veronica Aponte, and Marie-Anne
  Hudson.
\newblock {North American Breeding Bird Survey Dataset 1966 - 2021: U.S.
  Geological Survey data release}, 2022.
\newblock URL \url{https://doi.org/10.5066/P97WAZE5}.

\end{thebibliography}


\appendix 
\begin{center}
\end{center}
\setcounter{equation}{0}
\setcounter{figure}{0}
\setcounter{table}{0}
\setcounter{lemma}{0}
\setcounter{theorem}{0}
\setcounter{proposition}{0}
\setcounter{algorithm}{0}
\makeatletter
\renewcommand{\theequation}{A\arabic{equation}}
\renewcommand{\thetheorem}{A\arabic{theorem}}
\renewcommand{\thelemma}{A\arabic{lemma}}
\renewcommand{\theproposition}{A\arabic{proposition}}
\renewcommand{\thefigure}{A\arabic{figure}}
\renewcommand{\thetable}{A\arabic{table}}
\renewcommand{\thealgorithm}{A\arabic{algorithm}}

\section{Distribution theory}\label{sec:posterior_supp}
\subsection{Existence of improper prior on the discrepancy parameter}
Reparametrization of the discrepancy parameter $\mu = (\mu_1, \ldots, \mu_n)^\T$ into $q = - Q^\T D(\theta) \Tilde{\mu}$, where $\tilde{\mu} = (\mu^\T, 0_n^\T, \mu_\beta^\T L_\beta^{-\T}, 0_{nr}^\T)^\T$ is essential for obtaining the $\GCM$ joint prior distribution on the model parameters, as outlined in \eqref{eq:model_concise}. Our contribution here lies in proving two important results, Lemma~\ref{lemma:rank_supp} and Theorem~\ref{thm:improper_prior_supp}, that justifies constructing the improper prior that has not been addressed hitherto \citep[][Theorem~3.1]{bradley2023lgp}. We assume familiarity with notations introduced in Section~\ref{sec:bayeshier} of the main article. For convenience, rewrite $H$ in the main article as $H = [(I_n : X_{n \times p} : \Tilde{X}_{n \times nr})^{\T}; L_\gamma^{-1}]^{\T}$, where $L_\gamma = \mathrm{blkdiag}(I_n, L_\beta, L_z)$.
\begin{lemma}\label{lemma:rank_supp}
    If $H = [(I_n : X_{n \times p} : \Tilde{X}_{n \times nr})^{\T}; L_\gamma^{-1}]^{\T}$ and $Q = [Q_1^{\T}: Q_2^{\T}]^{\T}$, where $Q$ is $(2n+p+nr) \times n$ with $Q_1$ being $n \times n$, such that the $n$ columns of $Q$ are the unit norm orthogonal eigenvectors of $P_H$, the orthogonal projector on the column space of $H$, then $\mathrm{rank}(Q_1) = n$.
\end{lemma}
\begin{proof}
    From our definitions, we note that $Q Q^{\T} = I_{2n+p+nr} - P_H$, $P_H = H(H^{\T}H)^{-1}H^{\T}$, $Q^{\T} H = 0$ and $\mathrm{rank}(Q) = n$. Define $H_1 = [I_n : X : \Tilde{X}]$. It follows from $Q^{\T} H = 0$ that, $Q_1^{\T} H_1 + Q_2^{\T} L_\gamma^{-1} = 0$ and subsequently, $Q_2 = - L_\gamma^\T H_1^{\T} Q_1$, implying that $\mathcal{R} (Q_2) \subseteq \mathcal{R} (Q_1)$ where $\mathcal{R}$ denotes row space of a matrix. Hence the rank of $Q$ must equal $\dim (\mathcal{R}(Q_1))$, the number of independent rows of $Q_1$. As $\mathrm{rank}(Q) = n$, $\mathrm{rank}(Q_1) = \dim (\mathcal{R}(Q_1)) = n$.
\end{proof}
\begin{theorem}\label{thm:improper_prior_supp}
    In the hierarchical model \eqref{eq:model_final}, assumption of a vague prior on the parameter $\mu$ leads to the improper prior on the parameter $q$, given by $p(q) \propto 1$.
\end{theorem}
\begin{proof}
    Recall that $q = - Q^{\T} D(\theta) \Tilde{\mu}$ where $\Tilde{\mu} = (\mu^{\T}, \mu_\gamma^{\T} L_\gamma^{-\T})^{\T}$ with $\mu_\gamma  = (0_n^{\T}, \mu_\beta^{\T}, 0_{nr}^{\T})^{\T}$, and $Q = [Q_1^\T : Q_2^\T]^\T$ is the $(2n+p+nr) \times n$ matrix as defined in Lemma~\ref{lemma:rank_supp}, where $Q_1$ is $n \times n$ and $Q_2$ is $(n+p+nr) \times n$. Recall $D_(\theta) = \mathrm{blkdiag}(I_n, D_\gamma(\theta))$, where $D_\gamma(\theta) = \mathrm{blkdiag}(\sigma^2_\xi I_n, \sigma^2_\beta I_p, D_z(\theta_z))$ with $\theta = \{ \sigma^2_\xi, \sigma^2_\beta, \theta_z\}$. Then, we have $q = - Q_1^\T \mu - Q_2^\T D_\gamma(\theta) L_\gamma^{-1} \mu_\gamma$.
    
    Consider the sequence of priors on the discrepancy parameter as $p_{1, k}(\mu) = \mathcal{N}(\mu \given 0_n, \tau_k I_n)$ for a real sequence $\{ \tau_k \}_{k \in \mathbb{N}}$ with $\tau_k > 0, \forall k \in \mathbb{N}$ such that $\lim_{k} \tau_k = +\infty$. By Lemma~\ref{lemma:rank_supp}, for any $k$, the prior on $q$ induced by $p_{1,k}$ has the density $p_{2, k}(q) = \mathcal{N}(q \given - Q_2^\T D_\gamma(\theta) L_\gamma^{-1} \mu_\gamma, \tau_k Q_1^\T Q_1)$. Hence, the improper density $p_2(q) \propto 1$ arises from $\lim_{k} p_{2,k}(q) = p_2(q)$ for any $q \in \mathbb{R}^n$.
\end{proof}

\subsection{Derivation of posterior distribution}
We derive the posterior distribution in \eqref{eq:posterior_final} corresponding to the hierarchical model specified by \eqref{eq:model_concise} in the main article. 
The following result adapts the spatial intercept model of Theorem~3.1 in \cite{bradley2023lgp} to our spatially-temporally varying coefficient model, and, reformulates in a way that completely avoids the complicated conditional $\GCM$ distribution used previously in the literature \cite{bradley2023lgp}, thus offering a comprehensive and simplified derivation of the posterior distribution \eqref{eq:posterior_final}, 

\begin{theorem}\label{thm:posterior_supp}
Suppose $\{ y_i : i = 1, \ldots, n \}$ denotes responses distributed from the natural exponential family, $x_i$ is $p \times 1$ predictor corresponding to the $p \times 1$ fixed effect $\beta$, $\tilde{x}_i$ is the $r \times 1$ predictor with $r \times 1$ random slope $z_i = (z_{i1}, \ldots, z_{ir})^{\T}$. Then, for the Bayesian hierarchical model
\begin{equation}\label{eq:model_supp}
\begin{split}
    y_i \given \beta, z_i, \xi_i, \mu_i & \sim \EF (x_i^\T \beta + \tilde{x}_i^\T z_i + \xi_i - \mu_i; b_i, \psi_y), \quad i = 1, \ldots, n \\
    (\gamma^\T, q^\T)^\T & \sim \GCM (0_{2n+p+nr}, V, \alpha, \kappa, D, \pi; \psi)\;,
\end{split}
\end{equation}
where $\gamma = (\xi^\T, \beta^\T, z^\T)^\T$, with fine-scale variation $\xi = (\xi_1, \ldots, \xi_n)^\T$, $z = (z_1^\T, \ldots, z_r^\T)^\T$, $z_j$ is $n \times 1$ for each $j = 1, \ldots, r$. The parameter $\theta = \{\sigma^2_\xi, \sigma^2_\beta, \theta_z\}$ denotes a collection of scale parameters with prior $\pi(\theta)$, and $(n + p + nr) \times (n + p + nr)$ matrix $D(\theta) = \mathrm{blkdiag}(I_n, \sigma_\xi I_n, \sigma^2_\beta I_p, D_z(\theta_z))$ for $nr \times nr$ invertible matrix $D_z(\theta)$. The $(2n + p + nr) \times (2n + p + nr)$ matrix $V^{-1} = [H : Q]$ is assumed known, where $(2n + p + nr) \times (n + p + nr)$ matrix $H = [(I, X, \Tilde{X})^\T, L_\gamma^{-\T}]^\T$ with $n \times nr$ matrix $\Tilde{X} = [\mathrm{diag}(\tilde{x}_1), \ldots, \mathrm{diag}(\tilde{x}_r)]$, and, $L_\gamma = \mathrm{blkdiag}(L_\xi, L_\beta, L_z)$, and, $(2n + p + nr) \times n$ matrix $Q$ is made up of $n$ unit norm orthonormal eigenvectors of $I - H(H^\T H)^{-1} H^\T$ corresponding to the eigenvalue 1. The discrepancy parameter $\mu$ is reparametrized as $q = -Q^\T D(\theta) \Tilde{\mu}$, where $\Tilde{\mu} = (\mu^\T, \mu_\gamma^\T L_\gamma^{-\T})^\T$ with $\mu_\gamma = (\mu_\xi^\T, \mu_\beta^\T, \mu_z^\T)^\T$ collecting the prior location parameters of $\gamma$. The prior shape parameter $\alpha = (\alpha_\epsilon^\T, \alpha_\xi^\T, \alpha_\beta^\T, \alpha_z^\T)^\T$, and scale parameter $\kappa = (\kappa_\epsilon^\T, \kappa_\xi^\T, \kappa_\beta^\T, \kappa_z^\T)^\T$ are uniquely determined by the log partition functions in $\psi(h) = (\psi_y(h_1)^\T, \psi_\xi(h_2)^\T, \psi_\beta(h_3)^\T, \psi_z(h_4)^\T)^\T$ for $h = (h_1, h_2, h_3, h_4) \in \mathbb{R}^{2n+p+nr}$ with $h_1, h_2 \in \mathbb{R}^n$, $h_3 \in \mathbb{R}^p$ and $h_4 \in \mathbb{R}^{nr}$, where all of the log partition functions operate element-wise on the arguments. Then, 
\begin{equation*}
  (\gamma^\T, q^\T)^\T \given y \propto \GCM (0_{2n+p+nr}, V, \alpha^*, \kappa^*, D, \pi; \psi) \;,
\end{equation*}
where the posterior shape parameter $\alpha^* = ((y + \alpha_\epsilon)^\T, \alpha_\xi^\T, \alpha_\beta^\T, \alpha_z^\T)^\T$, and, posterior scale parameter $\kappa^* = ((b + \kappa_\epsilon)^\T, \kappa_\xi^\T, \kappa_\beta^\T, \kappa_z^\T)^\T$.
\end{theorem}
\begin{proof}
The posterior distribution is derived easily by observing that 
\begin{equation*}
p((\gamma^\T, q^\T)^\T \given y) \propto p(y \given (\gamma^\T, q^\T)^\T) \times p((\gamma^\T, q^\T)^\T)\;.
\end{equation*}
Define $H_1 = [I : X : \Tilde{X}]$. The observed likelihood can be rewritten as
\begin{equation}\label{eq:lik_supp}
\begin{split}
    p(y \given \gamma, \mu) & \propto \exp \left\{ y^\T (X\beta + \Tilde{X} z + \xi - \mu) - b^T \psi_y (X\beta + \Tilde{X} z + \xi - \mu) \right\} \\
    & = \exp \left\{ y^\T \left( H_1 \gamma - \mu \right) - b^T \psi_y \left( H_1 \gamma - \mu \right) \right\} \;.
\end{split}
\end{equation}
Following the definition of the $\GCM$ distribution as in \eqref{eq:CMdens}, we write the joint prior as
\begin{equation}\label{eq:prior_supp}
\begin{split}
p((\gamma^\T, q^\T)^\T) 
& \propto \int \exp \left\{ \alpha^\T D(\theta)^{-1} \begin{bmatrix} H & Q \end{bmatrix} \begin{bmatrix} \gamma \\ q \end{bmatrix} - \kappa^\T \psi \left(D(\theta)^{-1} \begin{bmatrix} H & Q \end{bmatrix}\begin{bmatrix} \gamma \\ q \end{bmatrix}\right) \right\} \pi(\theta)\, d\theta \\
& = \int \exp \left\{ \alpha^\T (D(\theta)^{-1} H \gamma - \tilde{\mu}) - \kappa^\T \psi (D(\theta)^{-1} H \gamma - \tilde{\mu}) \right\} \pi(\theta) \, d\theta \\
& = \int \exp \left\{ (\alpha_\epsilon^\T, \alpha_\xi^\T, \alpha_\beta^\T, \alpha_z^\T) \left( D(\theta)^{-1} \begin{bmatrix} H_1\\ L_\gamma^{-1} \end{bmatrix} \gamma - \begin{bmatrix} \mu \\ L_\gamma^{-1} \mu_\gamma \end{bmatrix} \right) \right. \\
& \left. \qquad - (\kappa_\epsilon^\T, \kappa_\xi^\T, \kappa_\beta^\T, \kappa_z^\T)\, \psi \left( D(\theta)^{-1} \begin{bmatrix} H_1\\ L_\gamma^{-1} \end{bmatrix} \gamma - \begin{bmatrix} \mu \\ L_\gamma^{-1} \mu_\gamma \end{bmatrix} \right) \right\} \pi(\theta) \, d\theta \\
& = \int \exp \left\{ \alpha_\epsilon^\T (H_1 \gamma - \mu) - \kappa_\epsilon^\T \psi_y(H_1 \gamma - \mu) \right\}\\ 
& \qquad \times \exp \left\{ \alpha_\gamma^\T (D_\gamma(\theta)^{-1} L_\gamma^{-1} \gamma - L_\gamma^{-1} \mu_\gamma) \right.\\
& \left. \qquad - \kappa_\gamma^\T \psi_\gamma(D_\gamma(\theta)^{-1} L_\gamma^{-1} \gamma - L_\gamma^{-1} \mu_\gamma) \right\} \pi(\theta) \, d\theta \;,
\end{split}
\end{equation}
where $\alpha_\gamma = (\alpha_\xi^\T, \alpha_\beta^\T, \alpha_z^\T)^\T$, $\kappa_\gamma = (\kappa_\xi^\T, \kappa_\beta^\T, \kappa_z^\T)^\T$, and, $D_\gamma(\theta) = \mathrm{blkdiag}(\sigma_\xi I_n, D_\beta(\theta), D_z(\theta))$. The second equality above follows from the fact that $q = - Q^\T D(\theta) \tilde{\mu}$, which implies $\tilde{\mu} = - D(\theta)^{-1} Qq$ since $Q^\T Q = I$.
Multiply \eqref{eq:lik_supp} with \eqref{eq:prior_supp} to obtain
\begin{equation}\label{eq:post_supp}
\begin{split}
p((\gamma^\T, q^\T)^\T \given y, \theta) & \propto \int \exp \left\{ (y + \alpha_\epsilon)^\T (H_1 \gamma - \mu) - (b + \kappa_\epsilon)^\T \psi_y(H_1 \gamma - \mu) \right\} \\
& \qquad \times \exp \left\{ \alpha_\gamma^\T (D_\gamma(\theta)^{-1} L_\gamma^{-1} \gamma - L_\gamma^{-1} \mu_\gamma) \right. \\
& \left. \qquad - \kappa_\gamma^\T \psi_\gamma(D_\gamma(\theta)^{-1} L_\gamma^{-1} \gamma - L_\gamma^{-1} \mu_\gamma) \right\} \pi(\theta) \, d\theta \\
& \propto \GCM(0_{2n+p+nr}, V, \alpha^*, \kappa^*, D, \pi; \psi) \;,
\end{split}
\end{equation}
where $\alpha^* = ((y + \alpha_\epsilon)^\T, \alpha_\gamma^\T)^\T$, and, $\kappa^* = ((b + \kappa_\epsilon)^\T, \kappa_\gamma^\T)^\T$. The final step follows from comparing the last line of \eqref{eq:prior_supp} with the density of a $\GCM$ distribution, as given in the first line of \eqref{eq:prior_supp}.
\end{proof}
The main article considers a special case of Theorem~\ref{thm:posterior_supp}, with $L_\xi = I_n$ and the location parameters $\mu_\xi = 0_n$ and $\mu_z = 0_{nr}$. Furthermore, we assume $\psi_\xi(\cdot)$, $\psi_\beta(\cdot)$, and $\psi_z(\cdot)$ to be the log-partition functions corresponding to the Gaussian density. This implies that the shape parameters $\alpha_\xi = 0_n$, $\alpha_\beta = 0_n$, $\alpha_z = 0_{nr}$, and the scale parameters $\kappa_\xi = (1/2)1_n$, $\kappa_\beta = (1/2) 1_p$, and $\kappa_z = (1/2) 1_{nr}$.

\subsection{Equivalence with alternative construction}\label{sec:equivalence_gcm_conditional}
In this section, we establish Theorem~\ref{thm:posterior_supp} as an alternative formulation of the hierarchical model in \eqref{eq:model_final}. We essentially show the equivalence of \eqref{eq:model_final} and \eqref{eq:model_concise}, thus implying that the hierarchical model \eqref{eq:model_final} indeed yields the posterior distribution \eqref{eq:posterior_final}. Instead of a $\GCM$ prior jointly on $(\gamma^\T, q^\T)^\T$, model \eqref{eq:model_final} places a $\GCMc$ prior
\begin{equation}\label{eq:xi_supp}
\xi \given \beta, z, \mu, \theta \sim \GCMc (\Tilde{\mu}_\xi, H_\xi, \alpha_\xi^\ast, \kappa_\xi^\ast, D_\xi, \pi_\xi; \psi_\xi^\ast)
\end{equation}
where $H_\xi = [I_n : L_\xi^{-1}]^\T$ is $2n \times n$, location parameter $\Tilde{\mu}_\xi = ((\mu - X \beta - \Tilde{X} z)^\T, L_\xi^{-1} \mu_\xi^\T )^\T$, shape and scale parameters $\alpha_\xi^\ast = (\alpha_\epsilon^\T, \alpha_\xi^\T)^\T$ and $\kappa_\xi^\ast = (\kappa_\epsilon^\T,  \kappa_\xi^\T)^\T$, respectively. And, $\psi_\xi^\ast(h) = (\psi_y(h_1)^\T, \psi_\xi(h_2)^\T)^\T$, where $h = (h_1, h_2) \in \mathbb{R}^{2n}$ with $h_1, h_2 \in \mathbb{R}^n$. This yields
\begin{equation}
\begin{split}
p(\xi \given \beta, z, \mu) &\propto \exp \left\{ (\alpha_\epsilon^\T, \alpha_\xi^\T) \left( D_\xi(\theta)^{-1} \begin{bmatrix} I_n \\ L_\xi^{-1} \end{bmatrix} \xi - \begin{bmatrix} \mu - X\beta - \Tilde{X}z \\ L_\xi^{-1} \mu_\xi \end{bmatrix} \right) \right. \\
& \left. \qquad - (\kappa_\epsilon^\T, \kappa_\xi^\T) \, \psi_\xi^\ast \left( D_\xi(\theta)^{-1} \begin{bmatrix} I_n \\ L_\xi^{-1} \end{bmatrix} \xi - \begin{bmatrix} \mu - X\beta - \Tilde{X}z \\ L_\xi^{-1} \mu_\xi \end{bmatrix} \right) \right\} \\
& = \exp \left\{ (\alpha_\epsilon^\T, \alpha_\xi^\T) \left( D_\xi(\theta)^{-1} \begin{bmatrix} I_n & X & \Tilde{X} \\ L_\xi^{-1} & 0 & 0 \end{bmatrix} \xi - \begin{bmatrix} \mu \\ L_\xi^{-1} \mu_\xi \end{bmatrix} \right) \right. \\
& \left. \qquad - (\kappa_\epsilon^\T, \kappa_\xi^\T) \, \psi_\xi^\ast \left( D_\xi(\theta)^{-1} \begin{bmatrix} I_n & X & \Tilde{X} \\ L_\xi^{-1} & 0 & 0 \end{bmatrix} \xi - \begin{bmatrix} \mu \\ L_\xi^{-1} \mu_\xi \end{bmatrix} \right) \right\} \;.
\end{split}
\end{equation}
We assign the same priors on $\beta$ and $z$, that is given by
\begin{equation}\label{eq:prior_supp2}
\begin{split}
p(\beta \given \theta) & \propto \exp \{ \alpha_\beta^\T (D_\beta(\theta)^{-1} L_\beta^{-1} \beta - L_\beta^{-1} \mu_\beta) - \kappa_\beta^\T \psi_\beta(D_\beta(\theta)^{-1} L_\beta^{-1} \beta - L_\beta^{-1} \mu_\beta) \} \\
p(z \given \theta) & \propto \exp \{ \alpha_z^\T (D_z(\theta)^{-1} L_z^{-1} z - L_z^{-1} \mu_z) - \kappa_z^\T \psi_z(D_z(\theta)^{-1} L_z^{-1} z - L_z^{-1} \mu_z) \} \;.
\end{split}
\end{equation}
Combine \eqref{eq:xi_GCMc} with the priors of $\beta$ and $z$ and assign an improper prior on $q$ given by $p(q) \propto 1$. Then the joint prior of $(\gamma^\T, q^\T)^\T$ is same as that of in \eqref{eq:model_final}. This can be seen by writing
\begin{equation}\label{eq:xi_GCMc}
\begin{split}
p((\gamma^\T, q^\T)) & = \int p(\xi \given \beta, z, \mu, \theta) \times p(\beta \given \theta) \times p(z \given \theta) \times p(q) \times \pi(\theta) \, d\theta \\
& \propto \int \exp \left\{ (\alpha_\epsilon^\T, \alpha_\xi^\T, \alpha_\beta^\T, \alpha_z^\T) \left( D(\theta)^{-1} \begin{bmatrix} I_n & X & \Tilde{X} \\ L_\xi^{-1} & 0 & 0 \\ 0 & L_\beta^{-1} & 0 \\ 0 & 0 & L_z^{-1} \end{bmatrix} \begin{bmatrix} \xi \\ \beta \\ z \end{bmatrix} - \begin{bmatrix} \mu \\ L_\xi^{-1} \mu_\xi \\ L_\beta^{-1} \mu_\beta \\ L_z^{-1} \mu_z \end{bmatrix} \right) \right. \\
& \left. \qquad - (\kappa_\epsilon^\T, \kappa_\xi^\T, \kappa_\beta^\T, \kappa_z^\T) \, \psi \left( D(\theta)^{-1} \begin{bmatrix} I_n & X & \Tilde{X} \\ L_\xi^{-1} & 0 & 0 \\ 0 & L_\beta^{-1} & 0 \\ 0 & 0 & L_z^{-1} \end{bmatrix} \begin{bmatrix} \xi \\ \beta \\ z \end{bmatrix} - \begin{bmatrix} \mu \\ L_\xi^{-1} \mu_\xi \\ L_\beta^{-1} \mu_\beta \\ L_z^{-1} \mu_z \end{bmatrix} \right) \right\} \\ 
& \qquad \times \pi(\theta)\, d\theta \\
& = \int \exp \{ \alpha^\T (D(\theta)^{-1} H \gamma - \tilde{\mu}) - \kappa^\T \, \psi (D(\theta)^{-1} H \gamma - \tilde{\mu}) \} \pi(\theta) \, d\theta\\
& = \int \exp \{ \alpha^\T D(\theta)^{-1} (H \gamma + Qq) - \kappa^\T \, \psi (D(\theta)^{-1} (H \gamma + Qq)) \} \pi(\theta) \, d\theta \\
& = \int \exp \{ \alpha^\T D(\theta)^{-1} V^{-1} (\gamma^\T, q^\T)^\T - \kappa^\T \, \psi (D(\theta)^{-1} V^{-1} (\gamma^\T, q^\T)^\T) \} \pi(\theta) \, d\theta\\
& \propto \GCM (0_{2n + p + nr}, V, \alpha, \kappa, D, \pi; \psi) \;,
\end{split}
\end{equation}
which aligns with our specification in \eqref{eq:model_final}, thus obtaining its equivalence with \eqref{eq:model_concise}.

\subsection{Multivariate spatial-temporal process}\label{subsec:mv_spt}
In this section, we outline some results related to the multivariate spatial-temporal process specification. Let $\calL = \{\ell_1,\allowbreak \ldots, \ell_n\}$ and $\Tilde{\calL} = \{ \ell_1, \ldots, \ell_{\tilde{n}} \}$ be two fixed sets of $n$ and $\tilde{n}$ distinct space-time coordinates in $\calD$ such that $\calL \cap \Tilde{\calL} = \emptyset$. Following \eqref{eq:multi_z}, define a $r \times 1$ multivariate spatial-temporal Gaussian process $z(\ell) \sim \GP (0, R(\cdot, \cdot; \thetasp) \Sigma)$ for some correlation kernel $R(\cdot, \cdot; \thetasp)$. Let $n \times r$ matrix $Z = [z_1 : \ldots : z_r]$ and $\tilde{n} \times r$ matrix $\Tilde{Z} = [\tilde{z}_1 : \ldots : \tilde{z}_r]$ denote the spatial-temporal process at $\calL$ and $\Tilde{\calL}$ respectively. Assume $\thetasp$ to be fixed.
\begin{theorem}\label{thm:matrix_t}
Suppose $[Z^\T, \Tilde{Z}^\T]^\T \given \Sigma \sim \MNorm_{n+\tilde{n}, r} (0, \Tilde{V}_z, \Sigma)$, where $\tilde{V}_z = [(R: C); (C^\T, \Tilde{R})]$ with $n \times n$ correlation matrix $R = R(\calL; \thetasp)$, $\Tilde{n}\times \Tilde{n}$ correlation matrix $\Tilde{R} = R(\Tilde{\calL}; \thetasp)$, and $C = [R(\ell, \ell'; \thetasp)]$ is $n \times \Tilde{n}$ with $\ell \in \calL, \ell' \in \Tilde{\calL}$, all assumed known under a fixed value of $\thetasp$. If $\Sigma \sim \IW(\nu_z + 2r, \Psi)$, then
\begin{itemize}
    \item[(a)] posterior distribution $\Sigma \given Z \sim \IW (n + \nu_z + 2r, Z^\T R^{-1} Z + \Psi)$,
    \item[(b)] marginal distribution $Z \sim \matrixT_{n, r} (\nu_z, 0, R, \Psi)$, and,
    \item[(c)] posterior predictive $\Tilde{Z} \given Z \sim \matrixT_{\tilde{n}, r} (\nu_z + n, C^\T R^{-1} Z, \Tilde{R} - C^\T R^{-1} C, Z^\T R^{-1} Z + \Psi)$.
\end{itemize}
\end{theorem}
\begin{proof}
We follow \cite{guptanagar_matrixvariate} for definitions of relevant matrix variate distributions used in the following proof. From the distribution of $[Z^\T, \Tilde{Z}^\T]^\T$ specified, the marginal distribution of $Z$ conditioned on $\Sigma$ is $Z \given \Sigma \sim \MNorm_{n, r}(0_{n,r}, R, \Sigma)$ which corresponds to the density
\begin{equation}\label{eq:zMN}
    p(Z \given \Sigma) = (2 \pi)^{-\frac{nr}{2}} |\Sigma|^{-\frac{n}{2}} |R|^{-\frac{r}{2}} \exp \left\{ -\frac{1}{2} \tr (\Sigma^{-1} Z^\T R^{-1} Z) \right\} \;,
\end{equation}
where $\tr(\cdot)$ denotes trace of a square matrix. Moreover, the prior $\Sigma \sim \IW(\nu_z + 2r, \Psi)$ has density
\begin{equation}\label{eq:SigmaIW}
    p(\Sigma) = \frac{|\Psi|^{\frac{1}{2}(\nu_z + r - 1)}}{2^{\frac{1}{2}r(\nu_z + r - 1)} \Gamma_r(\frac{1}{2}(\nu_z + r - 1))} |\Sigma|^{-\frac{1}{2} (\nu_z + 2r)} \exp \left\{ -\frac{1}{2} \tr (\Sigma^{-1} \Psi) \right\} \;,
\end{equation}
where $\Gamma_r(\cdot)$ denotes the multivariate gamma function \citep{guptanagar_matrixvariate}. From \eqref{eq:zMN} and \eqref{eq:SigmaIW}, we derive the posterior distribution $p(\Sigma \given Z)$ as
\begin{equation}\label{eq:postSigma}
\begin{split}
    p(\Sigma \given Z) & \propto p(Z \given \Sigma) \times p(\Sigma) \\
    & \propto |\Sigma|^{-\frac{n}{2}} \exp \left\{ -\frac{1}{2} \tr ( \Sigma^{-1} Z^\T R^{-1} Z) \right\} \times |\Sigma|^{-\frac{1}{2} (\nu_z + 2r)} \exp \left\{ -\frac{1}{2} \tr ( \Sigma^{-1} \Psi ) \right\} \\
    & = |\Sigma|^{-\frac{1}{2} (n + \nu_z + 2r)} \exp \left\{ -\frac{1}{2} \tr \left( \Sigma^{-1} (Z^\T R^{-1} Z + \Psi) \right) \right\} \\
    & \propto \IW \left( \Sigma \given n + \nu_z + 2r, Z^\T R^{-1} Z + \Psi \right) \;.
\end{split}
\end{equation}
The last step follows from comparing the terms up to a proportionality constant with that of a Inverse-Wishart density as given in \eqref{eq:SigmaIW}. This proves statement~(a) of Theorem~\ref{thm:matrix_t}.
In order to find the marginal distribution of $Z$, we use the following relation between the marginal, likelihood and the prior densities
\begin{equation*}
    p(Z) = \frac{p(Z \given \Sigma) \, p(\Sigma)}{p(\Sigma \given Z)}\;.
\end{equation*}
We use the posterior density $p(\Sigma \given Z)$ obtained above, and, evaluate
\begin{equation}
\begin{split}
    p(Z) &= (2 \pi)^{-\frac{nr}{2}} |\Sigma|^{-\frac{n}{2}} |R|^{-\frac{r}{2}} \exp \left\{ - \frac{1}{2} \tr (\Sigma^{-1} Z^\T R^{-1} Z) \right\} \\
    & \quad \times 2^{-\frac{1}{2}r(\nu_z + r - 1)} \Gamma_r \left(\frac{\nu_z + r - 1}{2} \right)^{-1} |\Psi|^{\frac{1}{2}(\nu_z + r - 1)} |\Sigma|^{-\frac{1}{2} (\nu_z + 2r)} \exp \left\{ -\frac{1}{2} \tr (\Sigma^{-1} \Psi) \right\} \\
    & \quad \times 2^{\frac{1}{2}r(n + \nu_z + r - 1)} \Gamma_r \left(\frac{n + \nu_z + r - 1}{2} \right) \lvert Z^\T R^{-1}Z + \Psi \rvert^{-\frac{1}{2}(n + \nu_z + r - 1)} \\
    & \quad \times |\Sigma|^{\frac{1}{2}(n + \nu_z + 2r)} \exp \left\{ \frac{1}{2} \tr \left( \Sigma^{-1} (Z^\T R^{-1} Z + \Psi) \right) \right\} \\
    & = (\pi)^{-\frac{nr}{2}} \frac{\Gamma_r(\frac{n + \nu_z + r - 1}{2})}{\Gamma_r (\frac{\nu_z + r - 1}{2})} |R|^{-\frac{r}{2}} |\Psi|^{\frac{1}{2}(\nu_z + r - 1)} \lvert Z^\T R^{-1}Z + \Psi \rvert^{-\frac{1}{2}(n + \nu_z + r - 1)}\\
    & = (\pi)^{-\frac{nr}{2}} \frac{\Gamma_r(\frac{n + \nu_z + r - 1}{2})}{\Gamma_r (\frac{\nu_z + r - 1}{2})} |R|^{-\frac{r}{2}} |\Psi|^{\frac{1}{2}(\nu_z + r - 1)} |\Psi|^{-\frac{1}{2}(\nu_z + n + r - 1)}\\ 
    & \quad \times \lvert I_r + \Psi^{-1} Z^\T R^{-1}Z  \rvert^{-\frac{1}{2}(\nu_z + n + r - 1)} \\
    & = (\pi)^{-\frac{nr}{2}} \frac{\Gamma_r(\frac{n + \nu_z + r - 1}{2})}{\Gamma_r (\frac{\nu_z + r - 1}{2})} |R|^{-\frac{r}{2}} |\Psi|^{-\frac{n}{2}} \lvert I_r + \Psi^{-1} Z^\T R^{-1}Z  \rvert^{-\frac{1}{2}(\nu_z + n + r - 1)} \\
    & = \matrixT_{n, r} \left( Z \given \nu_z, 0, R, \Sigma \right) \;,
\end{split}
\end{equation}
where $\matrixT_{n, r} ( Z \given \nu_z, 0, R, \Sigma )$ corresponds to the density of $n \times r$ matrix variate $t$-distribution with degrees of freedom $\nu_z$, location parameter $0_{n, r}$, $n \times n$ between-rows covariance matrix $R$, $r \times r$ between-columns covariance matrix, evaluated at $Z$. This completes the proof of statement~(b).

To prove the statement~(c), we proceed with joint distribution of $Z$ and $\Tilde{Z}$ conditioned on $\Sigma$. Consider the Cholesky decomposition of the random variable $\Sigma = L_\Sigma L_\Sigma^\T$. Conditioned on $\Sigma$, we transform the matrix variate $[Z^\T, \Tilde{Z}^\T]^\T$ by multiplying $L_\Sigma^{-\T}$ on the right. This yields the distribution of the transformed $(n + \tilde{n}) \times r$ random matrix $[L_\Sigma^{-1}Z^\T, L_\Sigma^{-1}\Tilde{Z}^\T]^\T \sim \MNorm_{n+\tilde{n}, r}(0, \Tilde{V}_z, I_r)$. This follows from properties on linear transformation of matrix normal variates and the identity $L_\Sigma^{-1} \Sigma L_\Sigma^{-T} = L_\Sigma^{-1} L_\Sigma L_\Sigma^{\T} L_\Sigma^{-T} = I_r$. This implies that this transformation has made the columns to be independent. We use the result on conditional multivariate Gaussian distribution independently on the columns of the transformed matrix random variable and then combine them, to obtain
\begin{equation}
\begin{split}
    \Tilde{Z} L_\Sigma^{-\T} \given Z, \Sigma &\sim \MNorm_{\tilde{n}, r} \left( C^\T R^{-1} Z L_\Sigma^{-\T}, \Tilde{R} - C^\T R^{-1} C, I_r \right)\\
    \iff \Tilde{Z} \given Z, \Sigma &\sim \MNorm_{\tilde{n}, r} \left( C^\T R^{-1} Z, \Tilde{R} - C^\T R^{-1} C, \Sigma \right) \;,
\end{split}
\end{equation}
which follows from multiplying by $L_\Sigma^{\T}$ on the right which transforms the location to $C^\T R^{-1} Z = C^\T R^{-1} Z L_\Sigma^{-\T} L_\Sigma^\T$, and changes the between-row covariance matrix to $L_\Sigma I_r L_\Sigma^{\T} = \Sigma$. Now, in order to obtain the conditional distribution $p(\Tilde{Z} \given Z)$, we need to integrate out $\Sigma$ as follows
\begin{equation*}
    p(\Tilde{Z} \given Z) = \int p(\Tilde{Z} \given Z, \Sigma)\, p(\Sigma \given Z) \, d\Sigma \;.
\end{equation*}
From (a), we know that the posterior distribution $\Sigma \given Z \sim \IW(n + \nu_z + 2r, Z^\T R^{-1} Z + \Psi)$. We use the result of the same integral in (b), and plug in $n + \nu_z$ in place of $\nu_z$, and $Z^\T R^{-1} Z + \Psi$ in place of $\Psi$, to obtain the marginal distribution $p(\Tilde{Z} \given Z)$ as a matrix variate $t$-distribution
\begin{equation*}
    \tilde{Z} \given Z \sim \matrixT_{\tilde{n}, r}(n + \nu_z, C^\T R^{-1} Z, \Tilde{R} - C^\T R^{-1} C, Z^\T R^{-1} Z + \Psi) \;.
\end{equation*}
This completes the proof of the statement~(c) of Theorem~\ref{thm:matrix_t}.
\end{proof}

\subsection{Posterior sampling of scale parameters}\label{subsec:recovery}
The projection expression \eqref{eq:proj1} in the main article is required to sample the fixed effects $\beta$ and the random effects $z$ from the marginal model \eqref{eq:model_concise}, obtained by integrating out the scale parameters $\sigma^2_\beta$ and $\theta_z$, respectively. Once we have obtained posterior samples $\{\beta^{(b)}, z^{(b)}\}$ from the posterior distribution \eqref{eq:posterior_final}, posterior samples of the scale parameters can be obtained easily. Note that, we need samples of $\sigma^2_\beta$ and $\theta_z$ from their marginal posterior $p(\sigma^2_\beta, \theta_z \given y, M)$ which can be rewritten as
\begin{equation}\label{eq:scale_posterior}
\begin{split}
    p(\sigma^2_\beta, \theta_z \given y, M) &= \int p(\sigma^2_\beta, \theta_z \given \beta, z, y, M) p(\beta, z \given y, M) d\beta dz \\
    & = \int p(\sigma^2_\beta, \theta_z \given \beta, z, M) p(\beta, z \given y, M) d\beta dz \\
    & = \int p(\sigma^2_\beta \given \beta, M) p(\theta_z \given z, M) p(\beta, z \given y, M) d\beta dz \;,
\end{split}
\end{equation}
where the penultimate step follows from the fact that both the scale parameters $\sigma^2_\beta$ and $\theta_z$ are conditionally independent of the data $y$, given $\beta$, $z$ and the fixed hyperparameters in $M$. The last step follows from the fact that $\sigma^2_\beta$ is independent of $\theta_z$ and $z$ given $\beta$ and $M$, which follows from the independent prior specification of $\beta$, $z$ and their corresponding scale parameters. Hence, given posterior samples $\{\beta^{(b)}, z^{(b)}\}$ from $p(\beta, z \given y, M)$, for each $\beta^{(b)}$, we need to sample $\sigma^{2(b)}_\beta \sim p(\sigma^2_\beta \given \beta^{(b)}, M)$, and, for each sample of $z^{(b)}$, we need to sample $\theta_z^{(b)} \sim p(\theta_z \given z^{(b)}, M)$. Thus, the posterior samples $\{ \beta^{b}, z^{(b)}, \sigma^{2(b)}_\beta, \theta_z^{(b)}\}$ will represent samples from the joint posterior distribution $p(\beta, z, \sigma^2_\beta, \theta_z \given y, M)$. Thus, choosing conditionally conjugate priors for the scale parameters (e.g., inverse-gamma/inverse-Wishart) plays a key role in conveniently sampling them from the joint posterior distribution.

Since, we place a multivariate Gaussian prior $\beta \given \sigma^2_\beta, M \sim \Norm (\mu_\beta, \sigma^2_\beta V_\beta)$ along with an inverse-gamma prior $\sigma^2_\beta \sim \IG (\nu_\beta/2, \nu_\beta/2)$, we derive the conditional posterior distribution of $\sigma^2_\beta$ as
\begin{equation}\label{eq:sigma_beta_posterior}
    p(\sigma^2_\beta \given \beta^{(b)}, M) = \IG \left(\sigma^2_\beta \biggiven \frac{\nu_\beta + p}{2}, \frac{\nu_\beta + Q_1(\beta^{(b)}; M)}{2} \right) \;,
\end{equation}
where $Q_1(v; M) = (v - \mu_\beta)^\T V_\beta^{-1} (v - \mu_\beta)$ for any $v \in \mathbb{R}^{p}$. Note that, the hyperparameters $\mu_\beta$ and $V_\beta$ are specified by $M$. Similarly, under the independent process model, we derive
\begin{equation*}
    p(\theta_z \given z^{(b)}, M) = \prod_{j = 1}^r p(\sigma^2_{z_j} \given z_j^{(b)}, M) \;,
\end{equation*}
where $\theta_z = \{\sigma^2_{z_j} : j = 1, \ldots, r\}$. This follows from the assumption of independent process $z_j(\ell)$ and independent prior on each $\sigma^2_{z_j}$ for $j = 1, \ldots, r$. Under assumption of the prior distribution $\sigma^2_{z_j} \sim \IG(\nu_{z_j}/2, \nu_{z_j}/2)$ for each $j$, we have
\begin{equation}\label{eq:z_scale_posterior1}
    p(\sigma^2_{z_j} \given z^{(b)}, M) = \IG \left(\sigma^2_{z_j} \biggiven \frac{\nu_{z_j} + n}{2}, \frac{\nu_{z_j} + Q_{2j}(z_j^{(b)}; M)}{2} \right), \quad j = 1, \ldots, r\;,
\end{equation}
where $Q_{2j}(v; M) = v^\T R(\thetasp_j) v$ for any $v \in \mathbb{R}^n$ and $j = 1, \ldots, r$. Here, the process parameters $\thetasp_j$ for each $j$ are specified by $M$. Under the multivariate process specification with inverse-Wishart prior $\Sigma \sim \IW (\nu_z + 2r, \Psi)$, we use Theorem~\ref{thm:matrix_t} in order to obtain $p(\theta_z \given z^{(b)}, M) = p(\Sigma \given z^{(b)}, M)$ as
\begin{equation}\label{eq:z_scale_posterior2}
    p(\Sigma \given z^{(b)}, M) = \IW \left(\Sigma \biggiven n + \nu_z + 2r, Q_3(z^{(b)}; M) \right) \;,
\end{equation}
where $Q_3(v; M) = \Psi + V^\T R(\thetasp)^{-1} V$, with $V \in \mathbb{R}^{n \times r}$ such that $\mathrm{vec}(V) = v$, for any $v \in \mathbb{R}^{nr}$. Here, the process parameters $\thetasp$ and the scale matrix $\Psi$ are specified by $M$.

\section{Computational details}\label{sec:sampling_supp}

In this section, we present an overview of the algorithms that we have developed to implement our proposed methodology in a computationally efficient way. We achieve this through optimizing linear algebraic operations by avoiding unnecessary expensive matrix operations that are slow and careful utilization of sparsity and other structural properties of matrices.

\subsection{Computationally efficient algorithm for projection}\label{sec:projalgo}
The projection expression \eqref{eq:proj1} is necessary to compute in order to obtain posterior samples of $\gamma = (\xi^\T, \beta^\T, z^\T)^\T$. Note that, it involves computing the inverse of the $(n + p + nr) \times (n + p + nr)$ matrix $H^\T H$, where $H = [(I_n : X : \Tilde{X}^\T, L_\gamma^{-\T})]^\T$. We proceed by first computing different submatrices of $(H^\T H)^{-1}$ and keeping it stored so that it can be reused whenever we need to find the projection for each $v^{(b)}$. We partition $H^\T H$ as
\begin{equation}
    H^\T H = \begin{bmatrix}
        2 I_n & X & \Tilde{X} \\
        X^\T & X^\T X + V_\beta^{-1} & X^\T \Tilde{X} \\
        \Tilde{X}^\T & \Tilde{X}^\T X & \Tilde{X}^\T \Tilde{X} + V_z^{-1}
    \end{bmatrix} = 
    \begin{bmatrix}
        A & B^\T \\ B & D
    \end{bmatrix}, \quad
    D = \begin{bmatrix}
        A_1 & B_1^\T \\ B_1 & D_1
    \end{bmatrix} \;,
\end{equation}
where $V_\beta = L_\beta L_\beta^\T$, $V_z = L_z L_z^\T$, $n \times n$ matrix $A = 2I_n$, $(p + nr) \times n$ matrix $B = [X : \Tilde{X}]^\T$, and $(p + nr) \times (p + nr)$ matrix $D$ is further partitioned into $p \times p$ matrix $A_1 = X^\T X + V_\beta^{-1}$, $nr \times p$ matrix $B_1 = \Tilde{X}^\T X$, and $nr \times nr$ matrix $D_1 = \Tilde{X}^\T \Tilde{X} + V_z^{-1}$. It suffices to develop the algorithm as computing $(H^\T H)^{-1} v$ for any $v \in \mathbb{R}^{n + p + nr}$. Rewrite $v$ as $v = (v_1^\T, v_2^\T)^\T$ where $v_1$ is $n \times 1$ and $v_2$ is $(p + nr) \times 1$. Then, $(H^\T H)^{-1} v$ can be computed using
\begin{equation}\label{eq:projalgo1}
    (H^\T H)^{-1} v = \begin{bmatrix}
        S_A^{-1} (v_1 - B^\T D^{-1} v_2) \\
        D^{-1} v_2 - D^{-1} B S_A^{-1} (v_1 - B^\T D^{-1} v_2)
    \end{bmatrix} \;,
\end{equation}
where $S_A = A - B^\T D^{-1} B$ denotes the Schur complement of the submatrix $A$ of $H^\T H$. Note that, computing $D^{-1}$ plays a central role in evaluating the projection \eqref{eq:projalgo1}. Similarly, we find $D^{-1} B$ as
\begin{equation}
    D^{-1}B = D^{-1} \begin{bmatrix} X^\T \\ \Tilde{X}^\T \end{bmatrix} = 
    \begin{bmatrix}
        S_{A_1}^{-1} (X - \Tilde{X} D_1^{-1} \Tilde{X}^\T X)^\T \\
        D_1^{-1} (\Tilde{X}^\T - \Tilde{X}^\T X S_{A_1}^{-1} (X - \Tilde{X} D_1^{-1} \Tilde{X}^\T X)^\T)
    \end{bmatrix}\;,
\end{equation}
where $S_{A_1} = A_1 - B_1^\T D_1^{-1} B_1$, $D_1 = \Tilde{X}^\T \Tilde{X} + V_z^{-1}$ and $B_1 = \Tilde{X}^\T X$. Here, the computational bottleneck lies in inverting $D_1$, which essentially involves two inversions of $nr \times nr$ matrices when approached naively i.e., $(\Tilde{X}^\T \Tilde{X} + V_z^{-1})^{-1}$. Here, we use the Sherman-Morrison-Woodbury identity
\begin{equation}\label{eq:sherman}
    D_1^{-1} = V_z - V_z \Tilde{X}^\T (I + \Tilde{X} V_z \Tilde{X}^\T)^{-1} \Tilde{X} V_z \;,
\end{equation}
which essentially boils down to inverting the $n \times n$ matrix $I + \Tilde{X} V_z \Tilde{X}^\T$, which is also known as the capacitance matrix in context of the Sherman-Morrison-Woodbury identity, thus avoiding finding Cholesky decomposition of a $nr \times nr$ matrix. The computation associated with evaluating the projection is dominated by the Cholesky factorization of the capacitance matrix which takes $O(n^3)$ flops. Moreover, computing $\Tilde{X} V_z \Tilde{X}^\T$ poses as a challenge since the matrix product $V_z \Tilde{X}^\T$ requires $O(rn^3)$ floating point operations (flops). We make use of both facts that $V_z$ is block-diagonal and $\Tilde{X}$ is sparse and have a particular structure, since $\Tilde{X} = [\mathrm{diag}(\Tilde{x_1}), \ldots, \mathrm{diag}(\Tilde{x}_r)]$. Once we have calculated $D_1^{-1}$, we successively find $D_1^{-1}B_1$, $S_{A_1}$, $D^{-1}B$ and $S_A^{-1}$. We compute these quantities only once, and use them for each iteration for sampling $\gamma^{(b)}$ for each $b$.


\subsection{Efficient Cholesky factor updates}
We utilize a block Givens rotation algorithm \citep[][Section 5.1.8]{GolubLoanMatrix4} for faster model evaluation during cross-validation. Our algorithm can be regarded as a block-level variant of \cite{kimcox2002_CV}. Following the setup preceding Algorithm~\ref{algo:stacking}, suppose the $k$th block $\chi_{[k]}$ for each $k$ contains adjacent indices of the data $\chi$. Moreover, suppose evaluating a model needs Cholesky decomposition of a $n \times n$ positive-definite matrix $R$, where $n$ is the number of observations in $\chi$. In the context of \eqref{eq:model_final}, $R$ may correspond to the correlation matrix of a spatial-temporal process. Suppose we have evaluated the model on the full data, and we have stored $L$, the upper triangular Cholesky factor of $R$. The consecutive indices of the partitioned data creates $K^2$ submatrices partitioning $R$ and its upper triangular Cholesky factor $L$ as,
\begin{equation*}
    R = \begin{bmatrix}
        R_{11} & R_{21}^\T & \cdots & R_{K1}^\T \\
        R_{21} & R_{22} & \cdots & R_{K2}^\T \\
        \vdots & \vdots & \ddots & \vdots \\
        R_{K1} & R_{K2} & \cdots & R_{KK}
    \end{bmatrix}, \quad
    L = \begin{bmatrix}
        L_{11} & L_{12} & \cdots & L_{1K} \\
        0 & L_{22} & \cdots & L_{2K} \\
        \vdots & \vdots & \ddots & \vdots \\
        0 & 0 & \cdots & L_{KK}
    \end{bmatrix}.
\end{equation*}
For each $k$ such that $1 < k < K$, consider the following representation of the matrices $R$, $L$, $R_{-k}$, the matrix corresponding to $\chi_{[-k]}$ and its upper triangular Cholesky factor $L_{-k}$.
\begin{align*}
    R = \begin{bmatrix}
        R_{11}^k & R_{21}^{k\T} & R_{31}^{k\T} \\
        R_{21}^k & R_{22}^k & R_{32}^{k\T} \\
        R_{31}^k & R_{32}^k & R_{33}^k
    \end{bmatrix},& \quad 
    L = \begin{bmatrix}
        L_{11}^k & L_{12}^k & L_{13}^k \\
        0 & L_{22}^k & L_{23}^k \\
        0 & 0 & L_{33}^k
    \end{bmatrix},& \\
    R_{-k} = \begin{bmatrix}
        R_{11}^k & R_{31}^{k\T} \\
        R_{31}^k & R_{33}^k
    \end{bmatrix},& \quad 
    L_{-k} = \begin{bmatrix}
        C_{11}^k & C_{13}^{k} \\
        0 & C_{33}^k
    \end{bmatrix} \;,&
\end{align*}
where $R_{22}^k = R_{kk}$. Then, Algorithm~\ref{algo:cholesky} finds $L_{-k}$ for each $k = 1, \ldots, K$. In Algorithm~\ref{algo:cholesky}, note that, no computation is needed to find $L_{-K}$ and $C_{11}^k$, $C_{13}^k$ for $1 < k < K$.
\begin{algorithm}[t]
\caption{Fast Cholesky updates of row-deletion using block Givens rotation.}\label{algo:cholesky}
\begin{algorithmic}[1]
\State \textbf{Input}: $R$, $L$, $k$; full matrix, its upper-triangular Cholesky factor, block to be deleted
\State \textbf{Output}: $L_{-k}$, the Cholesky factor of $R_{-k}$, the matrix with $k$th block deleted
\Function{\texttt{choleskyCV}}{$R$, $L$, k}
\If{$k = 1$}
\State $L_{-k} \gets \texttt{chol}(R_{-k})$
\Else
\If{$1 < k < K$}
\State $C_{11}^k \gets L_{11}^k$ (pre-computed)
\State $C_{13}^k \gets L_{13}^k$ (pre-computed)
\State $C_{33}^k \gets \texttt{chol}(L_{33}^{k \T} L_{33}^k + L_{23}^{k \T} L_{23}^k) $
\Else
\State $L_{-k} = $ upper left $(K-1) \times (K-1)$ submatrices of $L$ (pre-computed)
\EndIf
\EndIf
\State \Return $L_{-k}$
\EndFunction
\end{algorithmic}
\end{algorithm}

It is worth remarking that the ``naive approach'' for obtaining $L_{-k}$ involves Cholesky decomposition of $R_{k}$ for each $k$, and, hence, requiring $K^{-2}(K-1)^3 n^3/3$ floating point operations (flops). On the other hand, the time complexity of Algorithm~\ref{algo:cholesky} is of order
\begin{equation*}
    \sum_{k = 1}^K \left(1 - \frac{k}{K} \right)^3 \frac{n^3}{3} = \frac{n^3}{3 K^3} \sum_{k=1}^{K-1} i^3 =  \frac{n^3}{3 K^3} \frac{K^2 (K-1)^2}{4} = \frac{(K-1)^2}{4 K} \frac{n^3}{3}\;,
\end{equation*}
where the penultimate step follows from the sum of cubes of natural numbers. Hence, we show that Algorithm~\ref{algo:cholesky} is theoretically $4(K-1)/K$ times faster than the naive approach. If $K = 10$ \citep{vehtariCV2002}, then Algorithm~\ref{algo:cholesky} theoretically offers approximately 72\% efficiency in time complexity over the naive approach. However, it must be noted that modern linear algebra libraries are highly vectorised and are often multithreaded. Hence, it is difficult to accurately translate these purely theoretical flop counts to actual wall clock time. Nevertheless, Algorithm~\ref{algo:cholesky} is more efficient than the naive approach.

\subsection{Stacking algorithm}
Lastly, we detail the algorithm for implementing our proposed stacking framework utilising the aforementioned algorithms. Algorithm~\ref{algo:stacking} outlines the steps required for estimating the spatially-temporally varying coefficients model \eqref{eq:model_final} in the main article. The inputs $X$ and $\Tilde{X}$ in Algorithm~\ref{algo:stacking} correspond to the design matrices appearing in \eqref{eq:model_final}, $\calL$ denotes spatial-temporal locations, $\calM = \{M_1, \ldots, M_G\}$ denotes the collection of $G$ candidate models,
$N_s$ denotes number of samples to be drawn from each of these $G$ posterior distributions, $S$ denotes the number of samples to be drawn for evaluating the leave-one-out predictive densities as described in \eqref{eq:loo-pd}, and $K$ denotes the number of folds for cross-validation required for fast evaluation of leave-one-out predictive densities. Following a random permutation of the entire dataset, we construct a partition using consecutive indices of the permuted data. This step significantly accelerates evaluation of Cholesky factors using block Givens rotation \citep[][]{GolubLoanMatrix4} required during the cross-validation step detailed in Algorithm~\ref{algo:cholesky}. After partitioning the data, which is denoted by $\chi = (y, X, \Tilde{X}, \calL)$ into $K$ blocks, each block is denoted by $\chi_{[k]}$ and the data with the $k$th block deleted is denoted by $\chi_{[-k]}$ for $k = 1, \ldots, K$. Algorithm~\ref{algo:stacking} is written in a richer and more general context and can be easily adapted to spatial, spatial-temporal, spatially and spatially-temporally varying coefficients models.
\begin{algorithm}[t]
\caption{Predictive stacking algorithm for spatial-temporal GLM}\label{algo:stacking}
\begin{algorithmic}[1]
\State \textbf{Input}: $y$, $X$, $\Tilde{X}$, $\calL$, $N_s$, $S$, $\calM$, $K$
\State \textbf{Output}: Posterior samples for each model in $\calM$, optimal stacking weights $\hat{w}$
\Function{\texttt{stvcGLMstack}}{$y, X, \Tilde{X}, \calL, N, B, \calM, K$}
\State Partition the data $\chi$ into $K$ blocks $\chi_{[1]}, \ldots, \chi_{[K]}$.
\For{$g = 1$ to $g = G$}\textbf{ in parallel}
\State Fit model $M_g$ on $\chi$ and obtain posterior samples $\{ \beta_{g}^{(b)}, z_{g}^{(b)} \}_{b = 1}^{N_s}$ using \eqref{eq:proj2}
\For{$k = 1$ to $k = K$}
\State Obtain updated Cholesky factor using \texttt{choleskyCV}$(\cdot, \cdot, k)$.
\State Fit model $M_g$ on $\chi_{[-k]}$ and obtain samples $\{ \beta_{k, g}^{(s)}, z_{k,g}^{(s)} \}_{s = 1}^S$ using \eqref{eq:proj2}
\For{$s = 1$ to $s = S$}
\State Use \eqref{eq:cond_t} to predict $\tilde{z}_{k,g}^{(s)}$ at $\calL_{[k]}$ from $z_{k, g}^{(s)}$
\EndFor
\State Use \eqref{eq:loo-pd} to find $ p(y(\ell_i) \given y_{-i}, M_g), \forall \ell_i \in \calL_{[k]}$
\EndFor
\EndFor
\State Optimise \eqref{eq:stack_optim} to obtain optimal stacking weights $\hat{w}$
\State \Return Optimal model weights $\hat{w}$, $\{ \beta_{g}^{(b)}, z_{g}^{(b)}\}_{b = 1}^{N_s}$ for $g = 1, \ldots, G$
\EndFunction
\end{algorithmic}
\end{algorithm}

\section{Additional details on simulation experiments}\label{sec:simulation_supp}
\subsection{Predictive accuracy}
We present some additional results from the simulation experiments carried out in Section~\ref{sec:simulation} of the main article. Analogous to the main article, we evaluate predictive performance of our proposed methods using four simulated datasets each of sample size $n$, with spatial coordinates sampled uniformly inside the unit square $[0, 1]^2$ and temporal coordinates sampled uniformly within $[0, 1]$. The simulated datasets are based on the natural parameter $\eta(\ell) = x(\ell)^\T \beta + x(\ell)^\T z(\ell)$. For the first two simulated datasets, we simulate the responses as binomial count data with responses $y(\ell) \sim \mathrm{Binomial}(m(\ell), \pi(\ell))$ with $m(\ell)$, the number of trials sampled independently from a Poisson distribution with mean 20 and probability of success $\pi(\ell) = \mathrm{ilogit}(\eta(\ell))$ with $\mathrm{ilogit}(t) = \exp(t)/(1+\exp(t)), t \in \mathbb{R}$ and $\beta = (1, -0.5)$. For the remaining two datasets, we simulate the responses as binary data with responses $y(\ell) \sim \mathrm{Bernoulli}(\pi(\ell))$ with probability of success $\pi(\ell)$ is as defined above. We set $\tilde{x}(\ell) = x(\ell)$, deeming all covariates to have spatially-temporally varying coefficients. Under the independent process assumption, $z_j(\ell) \sim \mathcal{GP}(0, \sigma^2_{z_j} R_j(\cdot, \cdot))$ for $j = 1, 2$ with $\sigma^2_{z_1} = 0.25$, $\sigma^2_{z_2} = 0.5$ and, $(\phi_{11}, \phi_{21}) = (0.5, 2)$ and, $(\phi_{12}, \phi_{22}) = (1, 4)$. For the multivariate case, we assume $\Sigma = [(2, 0.5)^\T, (0.5, 2)^\T]^\T$ and $\phi_1 = 1$ and $\phi_2 = 2$. The choice of $\beta$ in each simulated data is such that the generated data do not contain excessive zeros. The simulation experiments are conducted with $n$ varying from 200 to 600 with a randomly chosen holdout sample of size $n_h = 100$ over a set of coordinates $\calL_h$. We present the results in Tables~\ref{tab:mlpd2_supp}~and~\ref{tab:mlpd3_supp}.
\begin{table}[t]
\centering
\resizebox{0.99\textwidth}{!}{%
{\begin{tabular}{@{}lccccccc@{}}
\toprule
\multirow{2}{*}{True Process} & \multirow{2}{*}{Model} & \multirow{2}{*}{Method} & \multicolumn{5}{c}{Sample size ($n$)} \\ \cmidrule{4-8} 
&  & & $n = 100$   & $n = 200$  & $n = 300$  & $n = 400$ & $n = 500$ \\ \cmidrule{1-8}
\multirow{4}{*}{Multivariate} & \multirow{2}{*}{Multivariate}  & Stacking & -3.76  & -3.64   & -3.59   & -3.47  & -3.41   \\
& & MCMC &   -3.67  &  -3.55    &  -3.49    &  -3.39   &  -3.34 \\ 
& \multirow{2}{*}{Independent} & Stacking & -3.90  &  -3.86  &  -3.76  & -3.69  &  -3.56 \\
& & MCMC &  -3.79   &  -3.75  &  -3.68  &  -3.57  &  -3.49 \\ 
\cmidrule{1-8}
\multirow{4}{*}{Independent} & \multirow{2}{*}{Multivariate}  & Stacking & -3.68  & -3.61   & -3.53   & -3.47  & -3.39  \\
& & MCMC &   -3.78  &  -3.72   &  -3.69    &  -3.64   &  -3.62 \\ 
& \multirow{2}{*}{Independent} & Stacking & -3.79  &  -3.76  &  -3.72  & -3.52  &  -3.38\\
& & MCMC & -3.63   &  -3.56  &  -3.49  &  -3.38  &  -3.34 \\ 
\bottomrule
\end{tabular}%
}
}
\caption{Predictive performance of stacking and MCMC under correct and misspecified spatial-temporal models on binomial count data. All values correspond to mean log-pointwise predictive density (MLPD) at $n_h = 100$ held-out locations, based on 5 replications.}
\label{tab:mlpd2_supp}
\end{table}

\begin{table}[t]
\centering
\resizebox{0.99\textwidth}{!}{%
{\begin{tabular}{@{}lccccccc@{}}
\toprule
\multirow{2}{*}{True Process} & \multirow{2}{*}{Model} & \multirow{2}{*}{Method} & \multicolumn{5}{c}{Sample size ($n$)} \\ \cmidrule{4-8} 
&  & & $n = 100$   & $n = 200$  & $n = 300$  & $n = 400$ & $n = 500$ \\ \cmidrule{1-8}
\multirow{4}{*}{Multivariate} & \multirow{2}{*}{Multivariate}  & Stacking & -4.36  & -4.29   & -4.19   & -4.07  & -3.89   \\
& & MCMC & -4.16  & -4.11   & -4.06   & -3.96  & -3.81\\ 
& \multirow{2}{*}{Independent} & Stacking & -4.41  &  -4.36  &  -4.23  & -4.19  &  -4.01 \\
& & MCMC &  -4.29   &  -4.21  &  -4.14  &  -4.09  &  -3.94 \\ 
\cmidrule{1-8}
\multirow{4}{*}{Independent} & \multirow{2}{*}{Multivariate}  & Stacking & -4.38  & -4.31   & -4.22   & -4.11  & -4.04  \\
& & MCMC &   -4.24  &  -4.19   &  -4.13    &  -4.06  &  -3.99 \\ 
& \multirow{2}{*}{Independent} & Stacking & -4.41  & -4.35   & -4.27   & -4.16  & -4.11\\
& & MCMC & -4.26   &  -4.20  &  -4.11  &  -4.07  &  -3.99 \\ 
\bottomrule
\end{tabular}%
}
}
\caption{Predictive performance of stacking and MCMC under correct and misspecified spatial-temporal models on binary data. All values correspond to mean log-pointwise predictive density (MLPD) at $n_h = 100$ held-out locations, based on 5 replications.}
\label{tab:mlpd3_supp}
\end{table}

\subsection{Weak identifiability of process parameters}
\begin{figure}[t]
    \centering
    \includegraphics[width=0.6\linewidth]{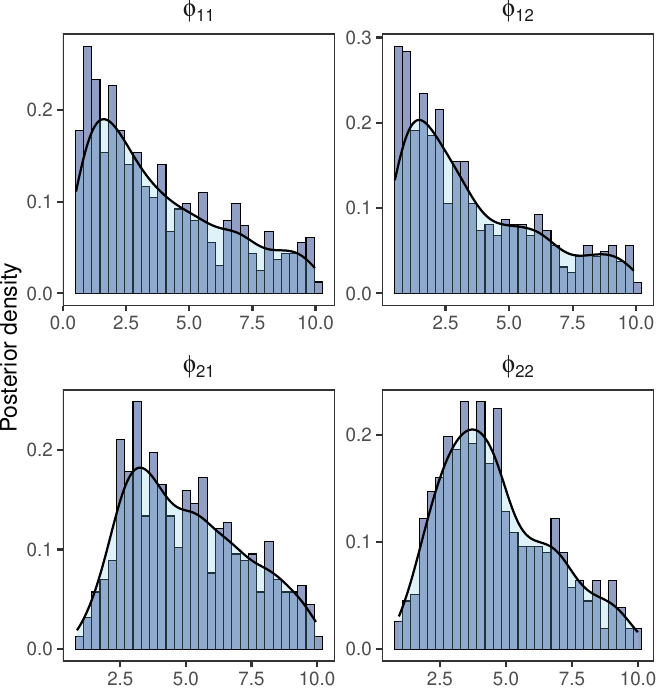}
    \caption{Histograms of posterior samples of the spatial-temporal process parameters obtained by MCMC on a fully Bayesian model under the independent process assumption on a simulated Poisson count data of sample 100.}
    \label{fig:plot_phi}
\end{figure}

Consider the simulated Poisson count data under the spatially-temporally varying coefficient model \eqref{eq:model_final} as described in Section~\ref{subsec:simulation}. Figure~\ref{fig:plot_phi} presents the posterior distributions of the process parameters that characterise the two spatial-temporal processes $z_1$ and $z_2$, corresponding to the intercept ($\phi_{11}$ and $\phi_{21}$) and the predictor ($\phi_{21}$ and $\phi_{22}$) respectively, obtained by MCMC (see Section~\ref{subsec:inference}). None of the process parameters in Fig.~\ref{fig:plot_phi} have concentrated around their corresponding true values, illustrating their weak identifiability. We also notice similarity in the posterior distributions of the temporal decay parameters $\phi_{11}$ and $\phi_{12}$ for both the processes $z_1$ and $z_2$ respectively. The same phenomenon is noticed in the posterior distributions of the spatial decay parameters $\phi_{21}$ and $\phi_{22}$. This between-process similarity in the posterior learning of the process parameters justifies the tenability of the ``reduced'' model that uses common spatial-temporal process parameters across all the $r$ processes in \eqref{eq:model_final}. This reduces the parameter space by $r$ folds. The reduced model further effectuates a substantial decrease in the number of candidate models required for the proposed stacking algorithm.

\begin{figure}[t]
    \centering
    \includegraphics[width=0.6\linewidth]{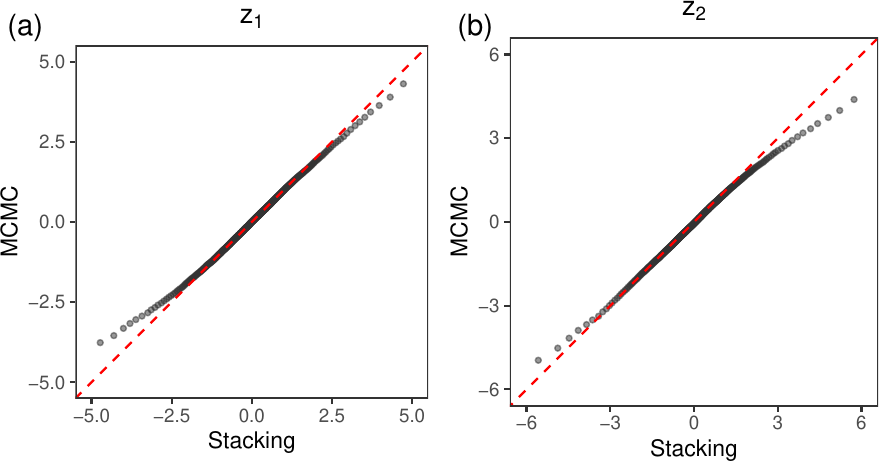}
    \caption{Plot of quantiles of the posterior distributions of the spatial-temporal random effects corresponding to (a) intercept and, (b) predictor obtained by stacking against MCMC with the $y=x$ reference as a red dashed line.}
    \label{fig:sptv_z_comb}
\end{figure}

Moreover, in addition to Fig.~\ref{fig:sptv_stack_vs_mcmc} in the main article, where we plot the posterior medians of the spatial-temporal random effects for the simulated Poisson count data, Figure~\ref{fig:sptv_z_comb} plots 100 quantiles of the combined posterior samples of spatial-temporal random effects obtained by stacking against quantiles obtained from MCMC. This demonstrates strong agreement between the posterior distributions obtained from the competing algorithms with most of the points concentrated along the$y=x$ line (red dashed).

\begin{figure}[t]
    \centering
    \includegraphics[width=0.9\linewidth]{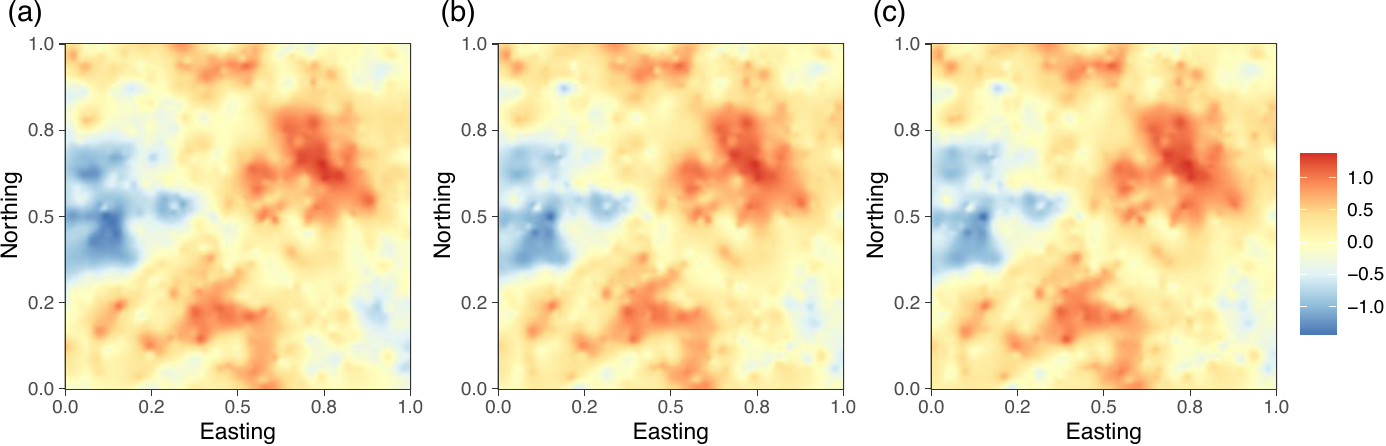}
    \caption{Comparison of interpolated surfaces of (a) the true spatial effects, with posterior median of spatial effects obtained by (b) MCMC, and, (c) our proposed predictive stacking algorithm on a simulated spatial count data.}
    \label{fig:spsurf}
\end{figure}

We also present results from an additional simulation experiment that demonstrates posterior learning of a random field modelling the underlying spatial or spatial-temporal processes. For convenient visualisation of the random field, we consider spatial data instead of spatial-temporal data in a continuous time domain. We simulate a dataset with responses distributed as $y(s) \sim \mathrm{Poisson}(\exp(x(s)^\T \beta + z(s))$ with sample size 1000. The locations are sampled uniformly inside the unit square $[0, 1]^2$. The explanatory vector $x(s)$ consists of an intercept and one predictor sampled from the standard normal distribution and the regression coefficients are taken as $\beta = (5, -0.5)$. The latent spatial process $z(s) \sim \GP(0, \sigma^2_z R(\cdot, \cdot; \phi, \nu))$ is a zero-centred Gaussian process with $\sigma^2_z = 0.4$, $\phi = 3.5$ and $\nu = 0.5$, where $R(\cdot, \cdot; \phi, \nu)$ is the Mat\'ern correlation function,
\begin{equation}\label{eq:matern}
    R(s, s'; \phi, \nu) = \frac{(\phi \lvert s - s' \rvert)^\nu}{2^{\nu - 1} \Gamma(\nu)} K_\nu (\phi \lvert s - s' \rvert)) \;,
\end{equation}
$\lvert s - s' \rvert$ is the Euclidean distance between $s$ and $s'$, and $\thetasp = \{\phi, \nu\}$. The parameter $\nu > 0$ controls the smoothness of the realised random field, $\phi$ is the spatial decay parameter, $\Gamma (\cdot)$ denotes the gamma function, and $K_\nu$ is the modified Bessel function of the second kind of order $\nu$ \citep[Chapter 10]{abramowitzstegun}. The model in \eqref{eq:model_final} can be modified for accommodating a spatial regression by considering $r = 1$ and subsequently $\Tilde{X} = I_n$. Here, $\calS$ is the domain of interest. We stack on the parameters $\Delta = \{\alpha_\epsilon, \sigma_\xi, \phi, \nu \}$ with hyperparameters $\nu_\beta = \nu_z = 3$. We also implement a fully Bayesian model with uniform priors $\mathrm{U}(0.5, 10)$ for $\phi$ and $\mathrm{U}(0.1, 2)$ for $\nu$. We modify our adaptive Metropolis-within-Gibbs algorithm accordingly. Figure~\ref{fig:spsurf} compares the posterior distributions of the spatial random effects obtained by predictive stacking and MCMC with its true values. We observe indistinguishable spatial surfaces of the posterior medians of the spatial random effects.

\subsection{Additional details on comparison with INLA}
\begin{figure}
    \centering
    \includegraphics[width=0.9\linewidth]{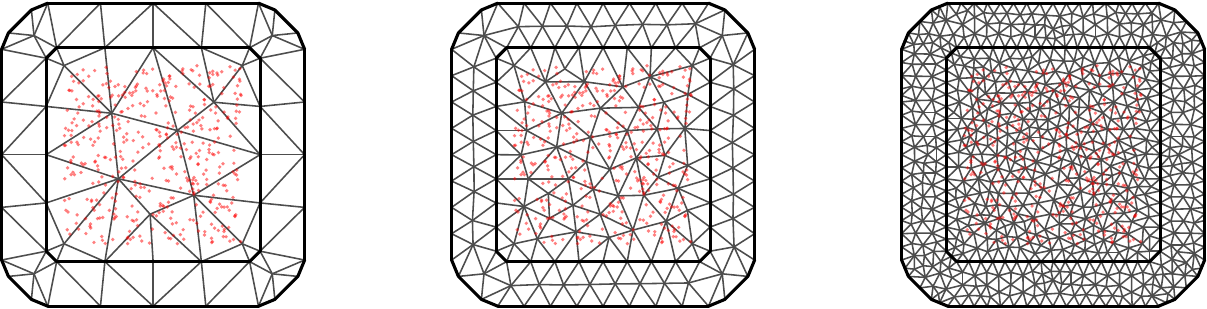}
    \caption{Spatial discretization at varying mesh resolutions constructed using INLA, with maximum edge lengths set to 0.5 (left), 0.25 (middle), and 0.15 (right). The 500 training and test locations are overlaid in red.}
    \label{fig:inla-mesh}
\end{figure}

Figure~\ref{fig:inla-mesh} illustrates the spatial discretization used in our INLA experiments, showing meshes constructed with varying maximum edge lengths $s_{\text{grid}} \in {0.5, 0.25, 0.15}$ and overlaid training and test locations (in red). 
We set the discrepancy parameter $\mu$ to zero, reducing the model to a traditional Poisson regression that is compatible with the INLA framework \cite{inla2009}. After constructing the spatial and temporal meshes, we specified the spatial–temporal processes and defined priors for their hyperparameters using the \textsf{stModel.define()} function from \textsf{INLAspacetime}. The SPDE projection matrix that links the latent process on the mesh to the observation locations was then constructed using \textsf{inla.spde.make.A()}, incorporating both observed and unobserved locations. We fitted the model through the \textsf{cgeneric} interface of \textsf{INLA}, and subsequently used \texttt{inla.posterior.sample()} to draw samples from the approximate posterior distribution of the latent process and hyperparameters. Using these samples together with the projection matrix, we recovered the posterior samples of the latent spatial–temporal process at the observed locations and generated predictions at held out sites. Finally, we computed the mean log pointwise predictive density (MLPD) at the held-out locations to assess predictive performance and compared total runtimes with those obtained from our predictive stacking approach.


\section{Additional data analysis}\label{sec:data_analysis_supp}
We offer some additional data analysis from the North American Breeding Bird Survey to supplement the results in Section~\ref{sec:data_analysis} of the main article. In the survey, the data are collected annually during the breeding season, primarily in June, along thousands of randomly established roadside survey routes in the United States and Canada. Routes are roughly 24.5~miles (39.2~km) long with counting locations placed at approximately half-mile (800~m) intervals, for a total of 50 stops. At each stop, a citizen scientist, highly skilled in avian identification, conducts a 3-minute point count recording all birds seen within a quarter-mile (400~m) radius and all birds heard. Routes are sampled once per year. In addition to avian count data, this dataset also contains route location information including country, state and the geographic coordinates of the route start point. The variable `car' records the total number of motorised vehicles passing a particular point count stop during the 3-minute count period. The variable `noise' represents the presence/absence of excessive noise defined as noise from sources other than vehicles passing the survey point (e.g. from streams, construction work, vehicles on nearby roads, etc.) lasting 45 seconds that significantly interferes with the observer's ability to hear birds at the stop during the 3-minute count period. More information on the survey can be found online at BBS 2022 data release (\url{https://www.usgs.gov/data/2022-release-north-american-breeding-bird-survey-dataset-1966-2021}).

\begin{figure}[t]
    \centering
    \includegraphics[width=0.9\linewidth]{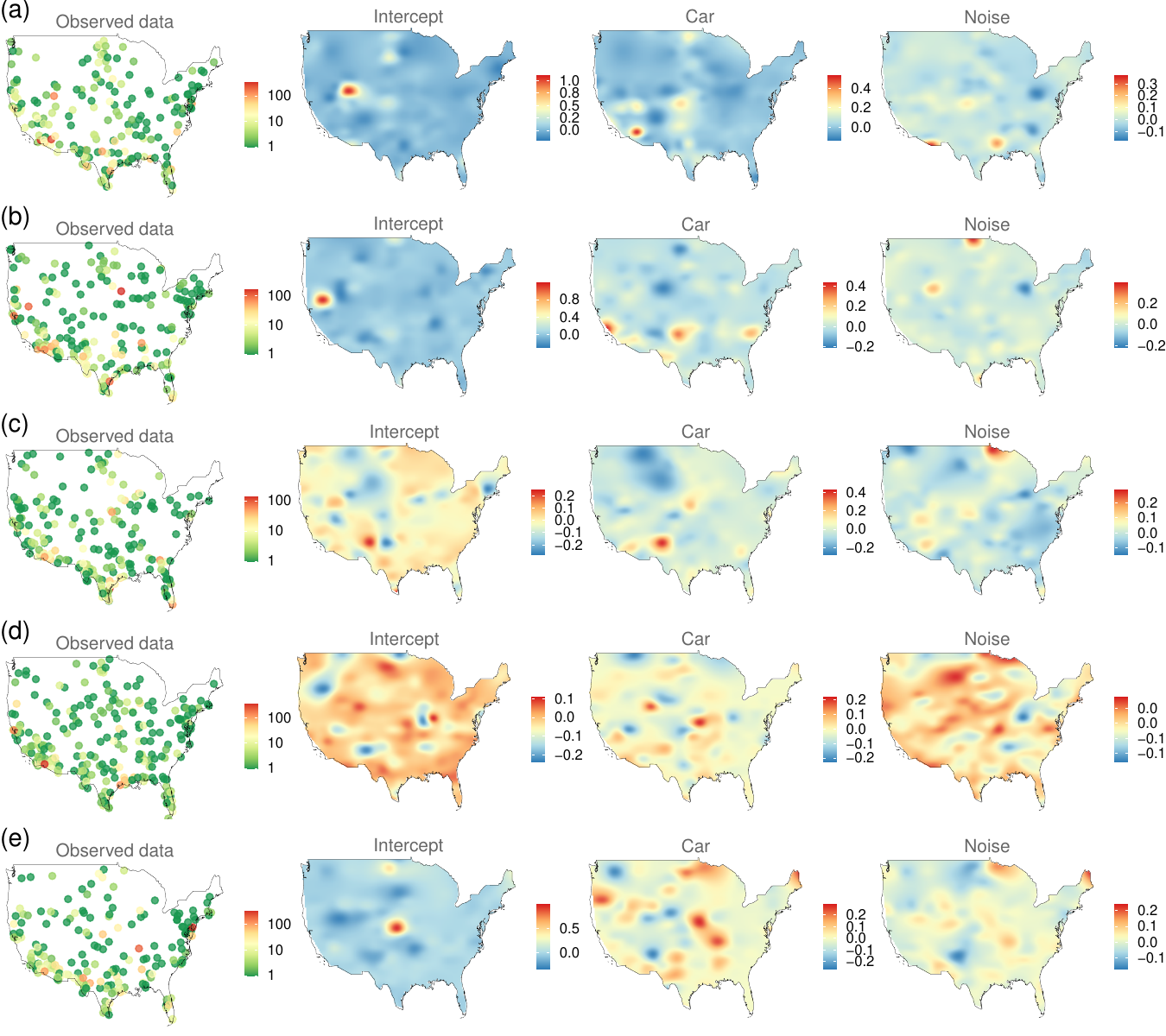}
    \caption{(a)--(e) Observed avian point count data and interpolated surfaces of posterior median of spatial-temporal random effects in the intercept as well as slopes of the variables `car' and `noise' for years 2010-2014.}
    \label{fig:bbs_supp1}
\end{figure}

\begin{figure}[t]
    \centering
    \includegraphics[width=0.9\linewidth]{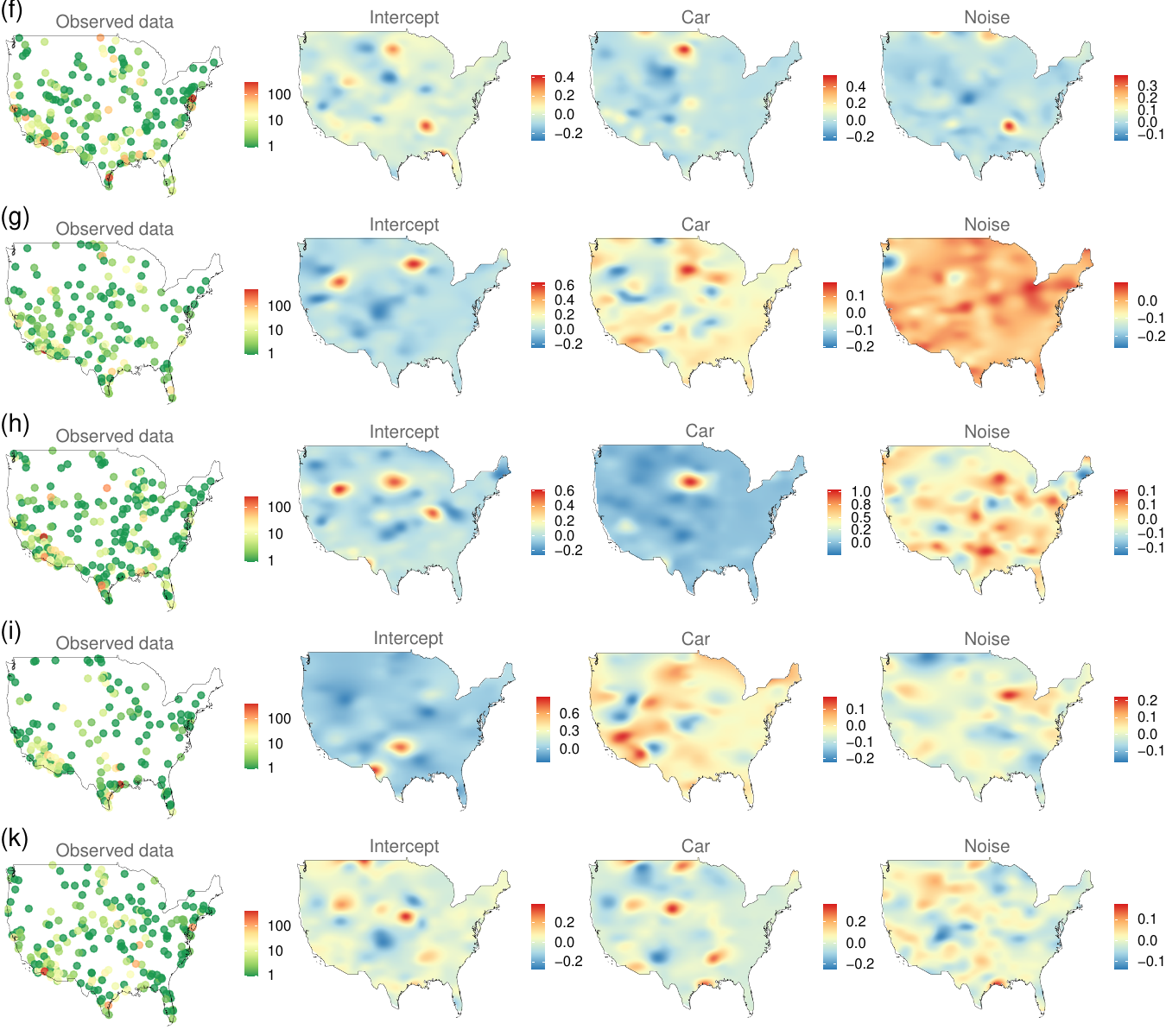}
    \caption{(f)--(k) North American Breeding Bird Survey (2015-19): Observed avian point count data and interpolated surfaces of posterior median of spatial-temporal random effects in the intercept as well as slopes of the variables `car' and `noise' for years 2015-2019.}
    \label{fig:bbs_supp2}
\end{figure}

The main article discusses the global regression coefficients under different models. In Table~\ref{tab:bbs_scale}, we present the posterior summary of the elements of the between-process covariance matrix under the multivariate process assumption.
\begin{table}[t]
\centering
\begin{tabular}{@{}lcccc@{}}
\toprule
 & Median & 2.5\%   & 97.5\% \\ \midrule 
$\sigma_{11}$ & 0.371 & 0.170 & 1.231 \\
$\sigma_{22}$ & 0.077 & 0.0428 & 0.214 \\
$\sigma_{33}$ & 0.134 & 0.351 & 1.397 \\
$\rho_{12}$ & -0.045 & -0.304 & 0.162 \\
$\rho_{13}$ & -0.487 & 0.197 & 0.731 \\
$\rho_{23}$ & -0.343 & -0.055 & 0.184 \\
\bottomrule
\end{tabular}
\caption{North American Breeding Bird Survey (2010-19): Posterior summary (95\% credible intervals) of elements of the between-process covariance matrix $\Sigma = ((\sigma_{ij}))$ with $\rho_{ij} = \sigma_{ij} / \sqrt{\sigma_{ii} \sigma_{jj}}$ denoting the correlation between the $i$th and $j$th processes in the varying coefficients. 2.5\% and 97.5\% represent quantiles.}
\label{tab:bbs_scale}
\end{table}
Figures~\ref{fig:bbs_supp1}~and~\ref{fig:bbs_supp2} display side-by-side plots of the observed point counts and interpolated spatial surfaces of the posterior median of the processes assigned to the intercept and the predictors `car' and `noise', revealing clear spatial-temporal patterns in the effect of the predictors.

\end{document}